\documentclass[%
 aip,
 amsmath,amssymb,
 reprint,%
]{revtex4-1}
\usepackage{graphicx}
\usepackage{subfigure}
\usepackage{dcolumn}
\usepackage{bm}
\usepackage{color}
\usepackage[T1]{fontenc}
\usepackage{mathptmx}
\usepackage{multirow}
\usepackage{array}
\usepackage{float}
\usepackage{etoolbox}
\usepackage{booktabs}
\usepackage{url}
\usepackage{CJKutf8}
\usepackage{pifont}
\usepackage{epstopdf, epsfig}
\usepackage{graphicx}
\usepackage{longtable}
\usepackage[
    colorlinks=true,
    linkcolor=blue,
    urlcolor=blue,
    citecolor=blue
]{hyperref}
\usepackage{placeins}
\usepackage{titlesec}
\titleclass{\subsubsubsection}{straight}[\subsubsection]
\newcounter{subsubsubsection}[subsubsection]
\renewcommand\thesubsubsubsection{\thesubsubsection.\arabic{subsubsubsection}}
\titleformat{\subsubsubsection}
  {\normalfont\normalsize\bfseries}{\thesubsubsubsection}{1em}{}
\titlespacing*{\subsubsubsection}
  {0pt}{1.5ex plus .1ex minus .2ex}{1ex plus .2ex}

\begin{document}
\preprint{AIP/123-QED}

\title{Enhanced oil recovery in reservoirs via diffusion-driven $\text{CO}_{2}$ flooding: Experimental insights and material balance modeling}


\author{Xiaoyi Zhang \begin{CJK*}{UTF8}{gbsn} (张晓祎) \end{CJK*}}
\affiliation{North China Institute of Aerospace Engineering, Langfang 065000, China}
\author{Rui Xu \begin{CJK*}{UTF8}{gbsn} (徐锐) \end{CJK*}}
\affiliation{Research Institute of Petroleum Exploration and Development, PetroChina Company Limited, Beijing 100083, China}
\author{Qing Zhao \begin{CJK*}{UTF8}{gbsn} (赵庆) \end{CJK*}}
\affiliation{No.2 Drilling Engineering Branch, PetroChina Bohai Drilling Engineering Co. LTD., Langfang 065007, China}
\author{Qian Cheng \begin{CJK*}{UTF8}{gbsn} (程倩) \end{CJK*}}
\affiliation{Beijing Deep Green Energy Technology Co. LTD., Beijing 100083, China}
\author{Rui Shen \begin{CJK*}{UTF8}{gbsn} (沈瑞) \end{CJK*}}
\thanks{Corresponding author: shenrui523@126.com}
\affiliation{Research Institute of Petroleum Exploration and Development, PetroChina Company Limited, Beijing 100083, China}
\author{Yanbiao Gan\begin{CJK*}{UTF8}{gbsn} (甘延标) \end{CJK*}}
 \thanks{Corresponding author: gan@nciae.edu.cn}
  \affiliation{North China Institute of Aerospace Engineering, Langfang 065000, China}

\date{\today}

\date{\today}

\begin{abstract}	
$\text{CO}_{2}$ flooding is central to carbon utilization technologies, yet conventional waterflooding models fail to capture the complex interactions between CO$_2$ and formation fluids. In this study, one- and two-dimensional nuclear magnetic resonance experiments reveal that $\text{CO}_{2}$ markedly enhances crude oil mobility during miscible displacement via multiple synergistic mechanisms, yielding a recovery factor of 60.97\%, which surpasses that of immiscible displacement (maximum $57.53\%$). Guided by these findings, we propose a convection–diffusion model that incorporates the diffusion coefficient ($D$) and porosity ($\phi$) as key parameters. This model captures the spatiotemporal evolution of the $\text{CO}_{2}$ front and addresses a key limitation of conventional formulations—the omission of diffusion effects. It improves predictions of gas breakthrough time and enables optimized injection design for low-permeability reservoirs. Extending classical material balance theory, we develop an enhanced $\text{CO}_{2}$ flooding equation that integrates critical transport phenomena. This formulation incorporates $\text{CO}_{2}$ diffusion, oil phase expansion, reservoir adsorption, and gas compressibility to describe the dynamic transport and mass compensation of injected $\text{CO}_{2}$. Validation through experimental and numerical data confirms the model's robustness and applicability under low-permeability conditions. The proposed framework overcomes limitations of physical experiments under extreme environments and offers theoretical insight into oil recovery enhancement and $\text{CO}_{2}$ injection strategy optimization.
\end{abstract}

\maketitle

\section{Introduction}\label{sec:I}

According to the latest Statistical Review of World Energy, fossil fuels account for approximately $81.5\%$ of global primary energy consumption, resulting in an estimated annual carbon dioxide ($\text{CO}_{2}$) emissions of nearly $40$ billion tons \cite{1Policies2025}.
Under the urgent need to address climate change, it is imperative to establish a multi-path collaborative carbon management technology system through carbon capture, utilization, and storage (CCUS) technologies. For example, injecting captured $\text{CO}_{2}$ into oil reservoirs for enhanced oil recovery ($\text{CO}_{2}$-EOR) can effectively reduce carbon emission intensity in industrial sectors.
Accordingly, leveraging $\text{CO}_{2}$ utilization within CCUS frameworks and investigating the underlying mechanisms of $\text{CO}_{2}$ flooding for improved oil recovery present significant technical opportunities and broad application potential.
With the depletion of conventional oil and gas resources, unconventional hydrocarbon development has become a strategic priority in the global energy transition.
Against this backdrop, $\text{CO}_{2}$ flooding, owing to its favorable geological adaptability, is increasingly replacing traditional methods such as horizontal well fracturing and water flooding. It is emerging as a core technology for improving oil recovery in unconventional reservoirs.

Based on extensive field trials and theoretical studies on $\text{CO}_{2}$ flooding, the primary displacement mechanisms and flow behaviors can be summarized as follows \cite{7, 8Study2021, 9, 10, 11, 12, 13, 14, 15, 16, 17, 18, 19, 20}
(1) Crude oil property modification: $\text{CO}_{2}$ dissolution reduces crude oil viscosity and induces oil swelling, enhancing its mobility.
(2) Interfacial tension reduction and extraction: $\text{CO}_{2}$ lowers the interfacial tension between the gas and oil phases, extracts light hydrocarbons, and facilitates dynamic miscibility at the displacement front.
(3) Mobility ratio improvement: $\text{CO}_{2}$ injection adjusts the mobility ratio between crude oil and formation water, improving sweep efficiency.
(4) Solution gas drive effect: Dissolved $\text{CO}_{2}$ generates a solution gas drive that promotes oil displacement.
(5) Molecular diffusion: The natural diffusivity of $\text{CO}_{2}$ enhances mass transfer across oil–gas interfaces.
(6) Carbonation reaction: $\text{CO}_{2}$ reacts with formation water and carbonate minerals (e.g., $\text{CaCO}_3 + \text{CO}_2 + \text{H}_2\text{O} \rightarrow \text{Ca}^{2+} + 2\text{HCO}_3^-$)
  increasing reservoir permeability by dissolving cementing materials.

Significant progress has been made by both domestic and international scholars in understanding the mechanisms and flow behavior of $\text{CO}_{2}$ flooding. These advancements can be categorized as follows:

\textbf{1. Theoretical modeling and steady-state flow:}
Badriyev \emph{et al.}~\cite{20} developed a two-dimensional mathematical model for incompressible fluid flow in homogeneous porous media under steady-state conditions. They derived analytical expressions for regional flow boundaries assuming a constant pressure gradient modulus.
Yi \emph{et al.}~\cite{21} applied unsteady-state seepage experiments and confirmed that water displacing gas follows Darcy's law under the tested conditions.

\textbf{2. $\text{CO}_{2}$-reservoir interactions:}
Narayanan \emph{et al.}~\cite{22} innovatively constructed a long-core physical model that accurately simulated reservoir alteration during $\text{CO}_{2}$ flooding, with strong agreement between experimental and numerical results.
Wang \emph{et al.}~\cite{23} systematically investigated the chemical alteration mechanisms induced by $\text{CO}_{2}$, elucidated the carbonation reaction process, and quantified the permeability changes in sandy conglomerates with varying clay mineral contents during supercritical $\text{CO}_{2}$ soaking.

\textbf{3. Multiphase flow and relative permeability:}
Zhang~\cite{24} determined relative permeability curves for the $\text{CO}_{2}$-oil system using unsteady-state displacement tests and analyzed how injection parameters affect gas–liquid flow behavior.
Dong~\cite{25} extended these experiments to assess the influence of injection pressure, rate, and volume on the relative permeability of $\text{CO}_{2}$-water and $\text{CO}_{2}$-oil systems.
Lun~\cite{26} showed that increasing displacement pressure significantly reduces oil–gas interfacial tension and enlarges the two-phase co-permeability zone, offering theoretical support for optimizing injection strategies.

\textbf{4. Modeling and numerical simulation:}
Jafari \emph{et al.}~\cite{27} proposed a four-component material balance model (oil, asphaltenes, light hydrocarbons, and water) based on multilayer adsorption theory to predict asphaltene behavior during $\text{CO}_{2}$ flooding.
Ampomah \emph{et al.}~\cite{28} utilized neural network algorithms to accurately forecast $\text{CO}_{2}$ storage efficiency and incremental oil recovery.
Zhu~\cite{29} used nuclear magnetic resonance (NMR) imaging to visualize flow evolution during near-miscible displacement and discovered that, at sufficiently high displacement pressures, the velocity ratio of the displacement front to injected volume stabilizes.

Existing models predominantly rely on ideal miscibility assumptions and neglect the impact of diffusion on displacement fronts, resulting in significant errors in $\text{CO}_{2}$ flooding recovery predictions.
And the $\text{CO}_{2}$ flooding process is governed by the interplay of multiple coupled mechanisms, making it challenging for any single model to fully capture the system's complexity. While significant progress has been achieved in modeling $\text{CO}{2}$ flooding under various pressure conditions, a unified and comprehensive theoretical framework remains elusive.

This study aims to leverage one-dimensional and two-dimensional nuclear magnetic resonance (NMR) experimental results, focusing on the viscosity reduction and extraction effects during $\text{CO}_{2}$ flooding, with emphasis on analyzing the underlying mechanisms under different displacement states. Further considering the diffusion effects in porous media flow, we quantify the coupling mechanism between diffusion and miscibility conditions, thereby addressing the limitations of existing models that overlook diffusion. Additionally, by integrating the diffusion coefficient ($D$) into a novel $\text{CO}_{2}$ flooding material balance equation and cross-validating experimental data with numerical simulations, this research provides new insights for advancing $\text{CO}_{2}$ flooding theory and optimizing operational processes.

\section{Experimental results and analysis of carbon dioxide flooding in Berea cores}\label{sec:II}

In this section, $\text{CO}_{2}$ flooding experiments were conducted under oil-saturated conditions using highly homogeneous Berea sandstone cores. A controlled-variable approach was adopted to systematically investigate the effect of injection pressure on $\text{CO}_{2}$ flooding efficiency. The use of standardized Berea cores effectively minimized permeability variability, ensuring reliable comparison across experimental runs.
Advanced characterization techniques were comprehensively applied, including one-dimensional nuclear magnetic resonance (1D NMR), two-dimensional $T_{1}-T_{2}$ NMR, and chromatographic analysis of crude oil components. These methods enabled a detailed investigation into the interaction mechanisms between $\text{CO}_{2}$ and crude oil. By varying the injected pore volume (PV), the influence of injection quantity on displacement performance was quantitatively assessed, providing experimental support for optimizing $\text{CO}_{2}$ flooding parameters.

\subsection{Experimental principles, equipment and samples}\label{sec:II-0}

$\text{CO}_{2}$ flooding technology is primarily categorized into two types: miscible flooding and immiscible flooding. The key distinction lies in whether the injected $\text{CO}_{2}$ can form a single phase with crude oil under specific reservoir conditions.
Immiscible displacement is suitable for low-to-medium permeability reservoirs ($1-50$ mD) where formation pressure is below the minimum miscibility pressure (MMP), typically $20-35$ MPa. It primarily improves oil mobility through physical effects induced by $\text{CO}_{2}$ dissolution, causing $15-35\%$ oil swelling and $30-80\%$ viscosity reduction, thereby enhancing microscopic displacement efficiency \cite{5Mechanism2020}.
Miscible displacement, however, requires stringent geological and engineering conditions:
Injection pressure must exceed MMP but remain below the reservoir fracture pressure threshold (typically $1.2$ MMP).
The reservoir must exhibit favorable porosity ($>15\%$) and permeability ($>10$ mD).\cite{6}
The key mechanism lies in the multiple-contact miscibility process $\text{CO}_{2}$ gradually extracts light ($\text{C}_{1}-\text{C}_{6}$), intermediate ($\text{C}_{7}-\text{C}_{15}$), and heavy ($>\text{C}_{16}$) hydrocarbon fractions from crude oil through successive contacts.
Under supercritical conditions ($T > 31.04^\circ \text{C}$, $P > 7.38$ MPa), a pseudo-single-phase flow forms, creating a fully miscible $\text{CO}_{2}$-oil displacement front at the interface, enabling more efficient oil displacement and higher recovery rates—though at significantly higher operational costs.

\begin{figure*}[!htbp]
\centering
\includegraphics[scale=1]{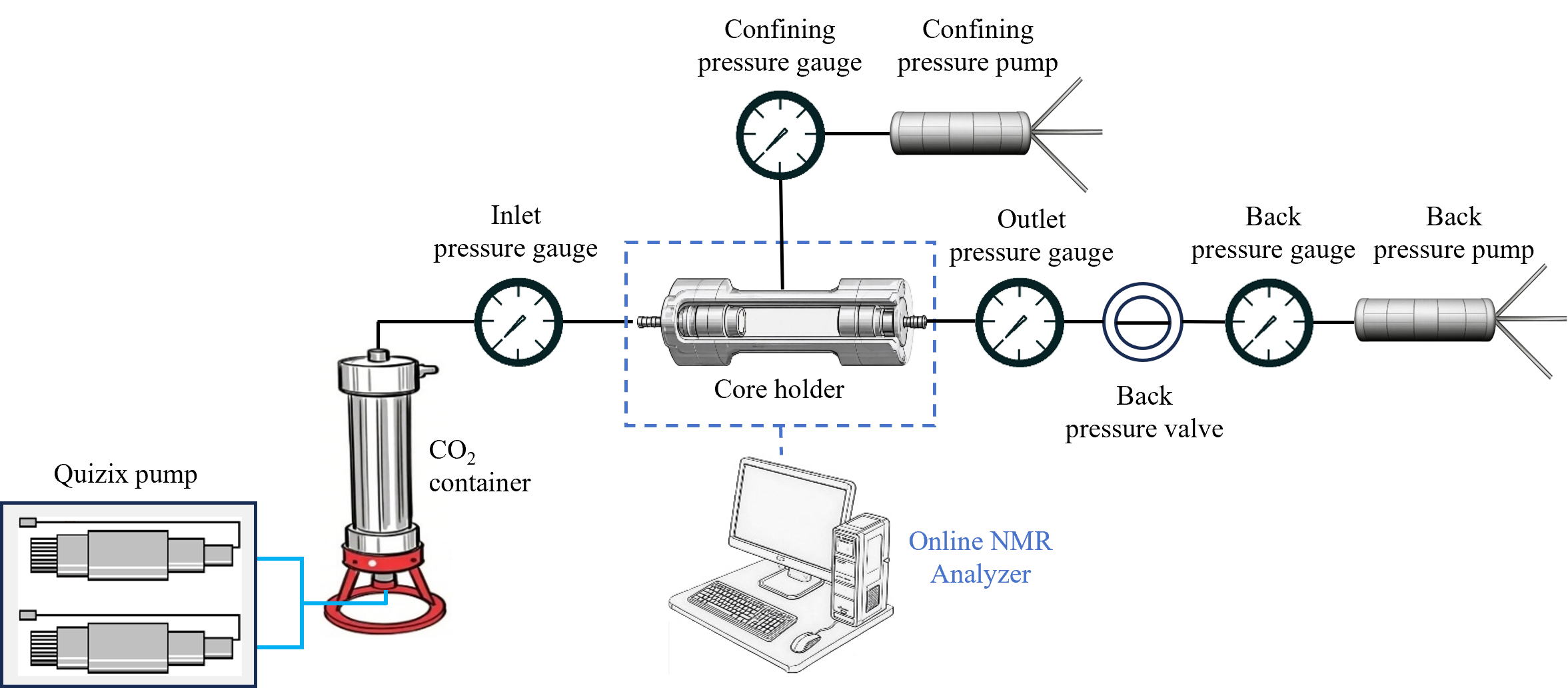}
\caption{Schematic diagram of the experimental setup for $\text{CO}_{2}$ flooding in saturated crude oil Berea core samples.}
\label{Fig0-1}
\end{figure*}

This study utilized high-homogeneity standard Berea cores to conduct $\text{CO}_{2}$ flooding experiments with oil-saturated cores, effectively eliminating the influence of core permeability variations on experimental results. Relevant core data are presented in Table \ref{table:table0}. During the experiments, the control variable method was systematically employed to investigate the impact of displacement pressure on $\text{CO}_{2}$ flooding efficiency.
\begin{table}[!htbp]
  \centering
  \caption{Reservoir $\text{CO}_{2}$ flooding natural cores data}
  \label{table:table0}
  \begin{tabular}{c|c|c|c|c|c}
    \hline
    Core ID & Lithology & Length & Diameter & Porosity & Permeability \\
    & & (cm) & (cm) & (\%) & (mD)  \\
    \hline
    Berea-1 & & 6.056 & 2.495 & 17.36 & 41.64  \\
    Berea-2 & Medium & 6.066 & 2.495 & 17.63 & 42.29  \\
    Berea-3 & permeability & 6.025 & 2.495 & 17.26 & 44.28  \\
    Berea-4 & sandstone & 6.061 & 2.495 & 17.34 & 48.14  \\
    Berea-5 &  & 6.084 & 2.495 & 17.45 & 44.84  \\
    \hline
  \end{tabular}
\end{table}
The NMR experiments on Berea cores were performed using a Reccore-04 core NMR analyzer, with measurement standards following SY/T 6490-2014 "Laboratory Measurement Specifications for NMR Parameters of Rock Samples". The $\text{CO}_{2}$ flooding experiments on oil-saturated Berea cores were completed using an SL-2018 $\text{CO}_{2}$ core experimental system, with the experimental procedure illustrated in Fig. \ref{Fig0-1}.

NMR $T_{1}-T_{2}$ mapping serves as a non-destructive technique for distinguishing hydrogen-containing components in sandstone. The color intensity in two-dimensional maps under the same scale enables qualitative analysis of shale oil fluid composition content and distribution within core samples. Free oil, adsorbed oil, and heavy components reside in organic pores, while bound water occupies inorganic pores. The transverse relaxation time ($T_{2}$) spectra of these fluids in rocks exhibit overlapping phenomena, making it difficult to effectively differentiate them using one-dimensional $T_{2}$ spectra alone.
Analysis of NMR characteristics of fluid components reveals significant differences in longitudinal relaxation times ($T_{1}$) among fluids in different rocks. Particularly, the $T_{1}/T_{2}$ ratio proves effective for evaluating oil-bearing information in sandstone. Experimental results demonstrate that various fluid components occupy distinct distribution ranges in $T_{1}-T_{2}$ two-dimensional spectra. Through data segmentation of $T_{1}-T_{2}$ spectra, detailed fluid component information can be obtained.
Based on this, existing research has proposed standard NMR spectra for different fluid components, indicating the existence of recognized two-dimensional NMR methods for oil-bearing detection. Furthermore, by varying the injected pore volumes (PV), this study quantitatively analyzes the influence of injection volume on $\text{CO}_{2}$ flooding efficiency, providing experimental basis for optimizing $\text{CO}_{2}$ flooding parameters.

This study follows the standard experimental procedure:

(1) The five Berea sandstone cores used in the experiment were vacuumed for $2$ hours, then saturated with kerosene. Subsequently, the five cores were placed in a pressure vessel and pressurized to $15.00$ MPa for additional kerosene saturation lasting over $12$ hours.

(2) The cores were then loaded into the $\text{CO}_{2}$ core flooding system. The system temperature was set to $97.3^\circ \text{C}$, and the cores were flooded with crude oil from Block Hei $79$ of Jilin Oilfield (whose minimum miscibility pressure of $22.10$ MPa was determined through slim-tube experiments) until achieving more than $5$ pore volumes (PV) of injection, ensuring complete crude oil saturation.

(3) The saturated cores were removed from the core holder, weighed and recorded, followed by one-dimensional NMR $T_{2}$ spectrum measurement and two-dimensional $T_{1}-T_{2}$ spectrum detection.

(4) The crude oil-saturated Berea cores were reassembled into the $\text{CO}_{2}$ flooding system. The inlet and outlet pressures were controlled separately using a high-pressure Quizix precision pump and back-pressure regulator, while the confining pressure was maintained at 23.11 MPa via a manual pump. Under constant net confining pressure, $\text{CO}_{2}$ flooding experiments were conducted through three injection stages with different PV numbers: $0.2$ PV, $0.4$ PV, and $2.0$ PV (cumulative $2.6$ PV).

(5) After each $\text{CO}_{2}$ injection stage, one-dimensional NMR $T_{2}$ spectra and two-dimensional $T_{1}-T_{2}$ spectra of the cores were measured.

(6) NMR $T_{2}$ and $T_{1}-T_{2}$ spectra were plotted for both pre-flooding and post-flooding states at different stages for comparative analysis.

(7) Portions of the produced oil were selected for NMR and compositional analysis to investigate $\text{CO}_{2}$ enhanced oil recovery mechanisms.

(8) Systematic $\text{CO}_{2}$ flooding experiments were subsequently performed on all five cores at designated pressures: $24.00$ MPa, $22.00$ MPa, $20.00$ MPa, $18.00$ MPa, and $16.00$ MPa.

\subsection{Experimental Study on carbon dioxide Flooding Mechanisms Using 1D NMR}\label{sec:II-1}

\begin{figure}[!htbp]
\centering
\includegraphics[scale=0.28]{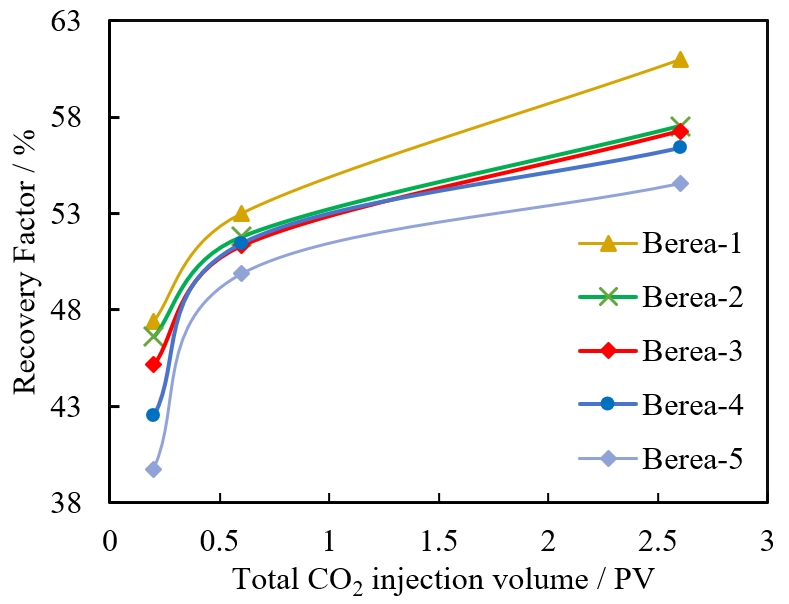}
\caption{Recovery factor as a function of total injected $\text{CO}_{2}$ PV for five Berea sandstone cores. The Berea-1 core, subjected to miscible flooding, exhibits a distinct upward deviation compared to the other cores. Notably, its ultimate recovery factor is 3.44\% higher than that of the Berea-2 core, while the difference between Berea-2 and Berea-5 is only 2.95\%.}
\label{Fig0}
\end{figure}

By systematically evaluating the displacement efficiency at different injection volumes, $\text{CO}_{2}$ flooding recovery data were obtained for five Berea cores under varying pressure conditions (see Table~\ref{table:table1}).
And Fig. \ref{Fig0} presents a line chart demonstrating the evolution of recovery factors in different core samples as $\text{CO}_{2}$ injection volume increases.

\begin{table}[!htbp]
  \centering
  \caption{Recovery factors of $\text{CO}_{2}$ flooding in five Berea sandstone cores under varying injection pressures.}
  \label{table:table1}
  \begin{tabular}{c|c|ccc}
    \hline
    \multirow{2}{*}{Core ID} & Pressure & \multicolumn{3}{c}{Recovery Factor (\%)} \\
    \cline{3-5}
    & (MPa) & 0.2 PV & 0.4 PV & 2.0 PV \\
    \hline
    Berea-1 & 24.00 & 47.40 & 52.99 & 60.97 \\
    Berea-2 & 22.00 & 46.62 & 51.80 & 57.53 \\
    Berea-3 & 20.00 & 45.16 & 51.32 & 57.26 \\
    Berea-4 & 18.00 & 42.52 & 51.46 & 56.41 \\
    Berea-5 & 16.00 & 39.72 & 49.88 & 54.58 \\
    \hline
  \end{tabular}
\end{table}

Comprehensive analysis of experimental data (Table~\ref{table:table1}) and intuitive comparison shown in Fig.\ref{Fig0} indicates that $\text{CO}_{2}$ flooding exhibits excellent displacement efficiency in low-permeability Berea sandstone cores, maintaining stable recovery across varying injection pressures. Mechanistic investigations suggest that $\text{CO}_{2}$ improves oil recovery through several synergistic mechanisms: (1) selective extraction of light hydrocarbons reduces crude oil viscosity; (2) the pronounced expansion behavior of supercritical $\text{CO}_{2}$ enables it to occupy additional pore space; and (3) the solution gas drive effect, especially under lower pressure conditions ($16-18$ MPa), further contributes to enhanced recovery performance.

During the cumulative injection of $0.6$ PV of $\text{CO}_{2}$ (including a secondary injection of $0.4$ PV), oil recovery continued to increase significantly, though at a slower rate than during the initial $0.2$ PV stage. This trend reveals two important insights:
(1) Substantial early-stage recovery can be achieved in low-permeability Berea cores even with small PV injection volumes, due to their favorable flow capacity and the rapid diffusion of $\text{CO}_{2}$.
(2) Further increasing the injected volume from $0.6$ PV to $2.6$ PV continues to enhance recovery, highlighting the importance of prolonged $\text{CO}_{2}$ contact time in low-permeability reservoir development.
These findings provide an experimental basis for optimizing $\text{CO}_{2}$ flooding strategies: using smaller PV volumes during early development to achieve rapid production, followed by higher PV injection to maintain stable output and maximize long-term recovery efficiency.

When the cumulative $\text{CO}_{2}$ injection volume reached $2.6$ PV (including a secondary injection of $2.0$ PV), all core samples achieved favorable ultimate recovery. This result confirms that large PV injection is a necessary condition for economically efficient development in low-permeability reservoirs using $\text{CO}_{2}$ flooding.
Based on these findings, it is recommended that field-scale implementations adopt a reservoir-specific injection strategy, whereby smaller PV volumes are used during the early development phase to achieve rapid production response, followed by larger PV injections in the later stages to ensure high ultimate recovery.

Based on the experimentally determined minimum miscibility pressure (MMP = $22.10$ MPa), results show that only the Berea-1 core (24.00 MPa) achieved miscible displacement conditions within the tested pressure range, exhibiting significantly higher ultimate recovery compared to the other samples. Table \ref{table:table1} and Fig. \ref{Fig0} show that its recovery factor is $3.44\%$ higher than that of Berea-2 ($22.00$ MPa). As a comparison, recovery factor of Berea-2 is only $0.27\%$ higher than that of Berea-3 ($20.00$ MPa) and even only $2.95\%$ higher than that of Berea-5 ($16.00$ MPa), which is still smaller than the gap between it and Berea-1. This discrepancy cannot be solely explained by pressure differences. It can be concluded that achieving miscible displacement conditions leads to significantly enhanced flooding efficiency.
The remaining four cores (Berea-2 to Berea-5, with pressures ranging from $16.00$ to $22.00$ MPa) operated under immiscible conditions but still demonstrated favorable displacement performance. Notably, oil recovery in the immiscible group displayed pressure sensitivity: as injection pressure decreased from $22.00$ MPa to $16.00$ MPa, the recovery factor declined correspondingly.
These findings confirm the superior efficiency of miscible displacement in enhancing oil recovery and underscore the importance of optimizing injection pressure even under immiscible conditions to ensure effective reservoir development.

\begin{figure}[!htbp]
\centering
\includegraphics[scale=0.37]{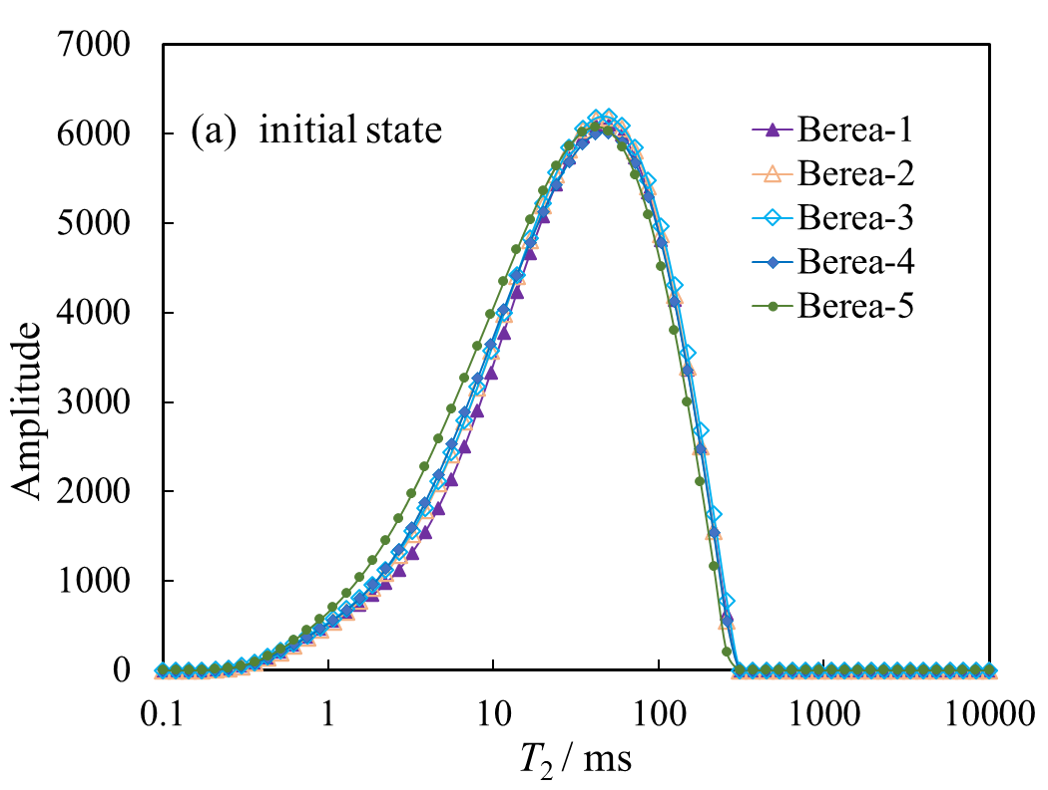}\\
\includegraphics[scale=0.37]{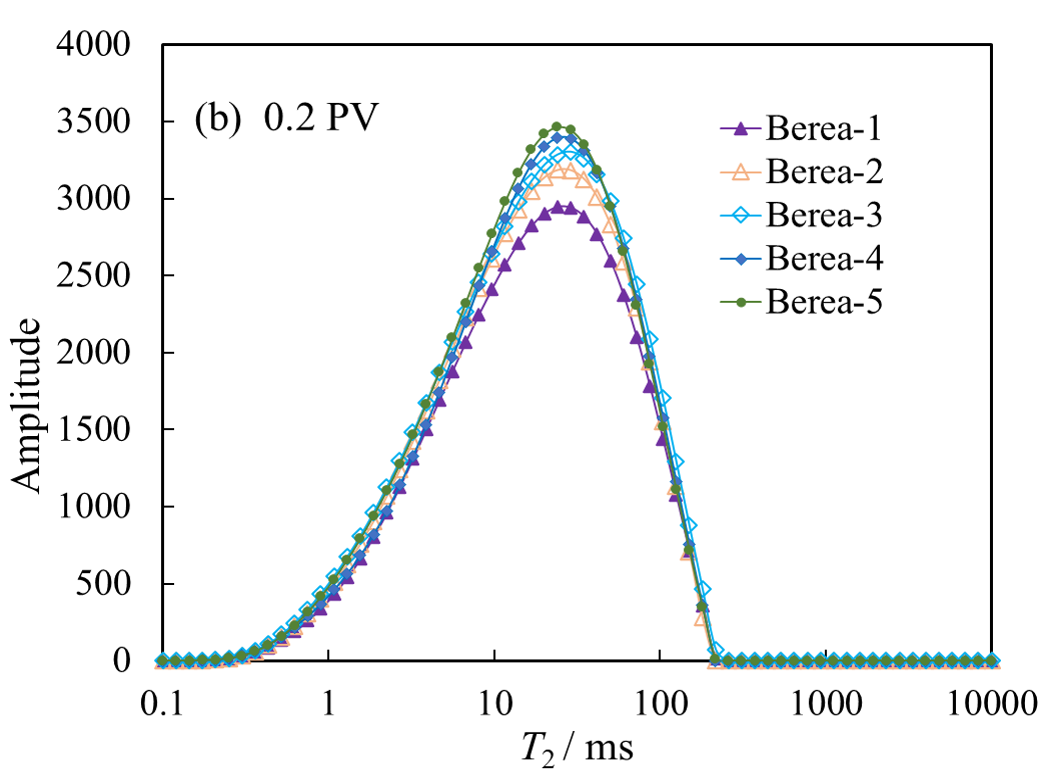}\\
\includegraphics[scale=0.37]{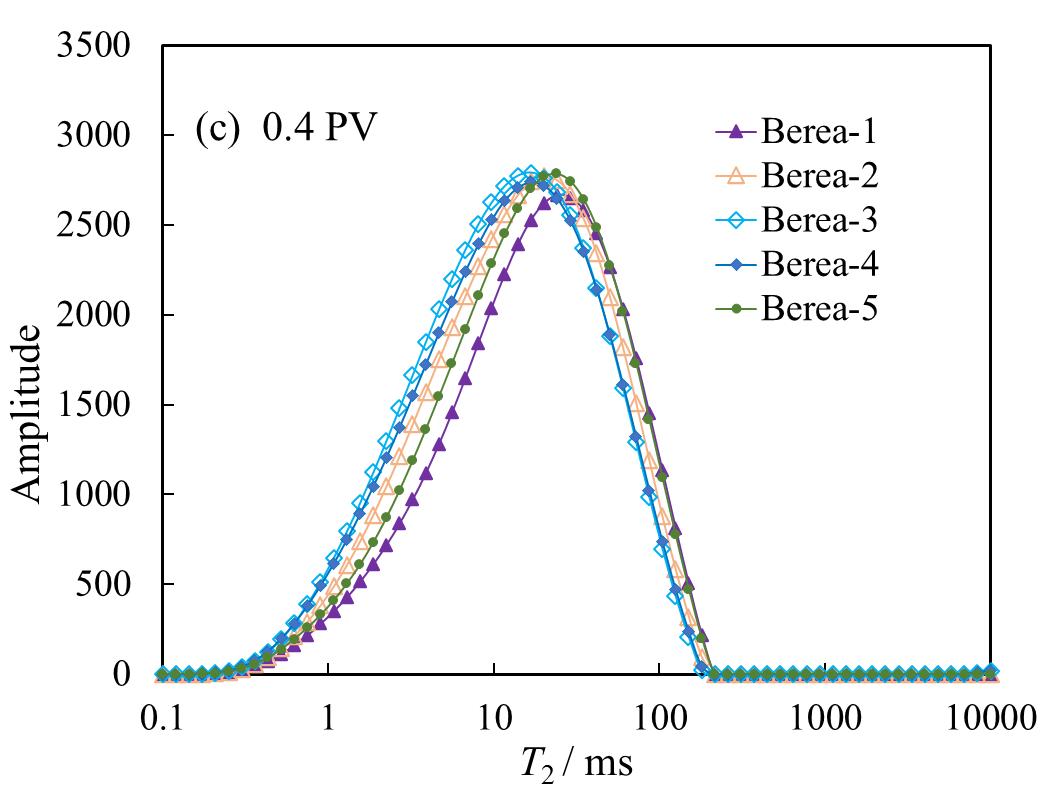}\\
\includegraphics[scale=0.37]{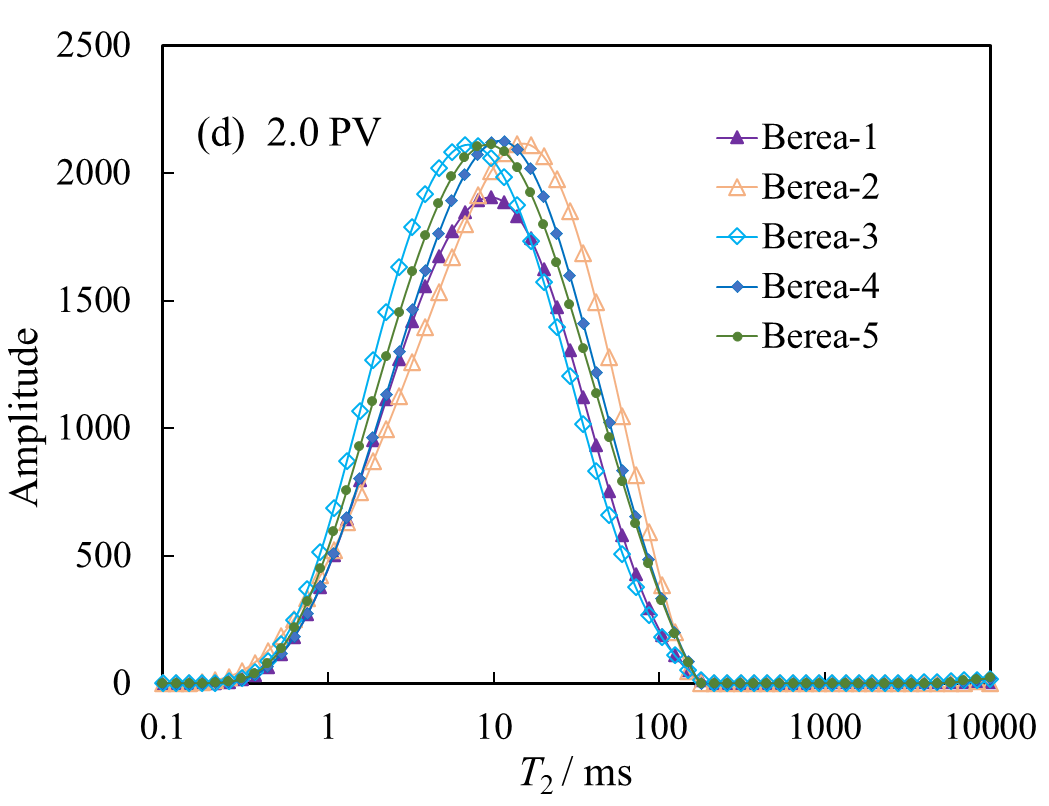}
\caption{Comparison of $T_{2}$ spectra at different displacement states for five Berea cores. Across all injection volumes, each core exhibits a single peak in the $T_2$ spectrum, indicating a highly uniform pore size distribution. The area under each peak corresponds to the crude oil volume retained in the core. As CO$_2$ flooding progresses, the residual oil content in the cores gradually decreases, with Berea-1 core showing the smallest remaining oil volume.}
\label{Fig1}
\end{figure}

Figure \ref{Fig1} presents the comparative $T_{2}$ spectra of five Berea sandstone cores under four distinct displacement conditions:
(1) the initial state saturated with crude oil;
(2) after injection of $0.2$ PV $\text{CO}_{2}$ (cumulative $0.2$ PV);
(3) after an additional $0.4$ PV injection (cumulative $0.6$ PV);
(4) after a final $2.0$ PV injection (cumulative $2.6$ PV).

As shown in Fig.~\ref{Fig1}, the $T_{2}$ spectra of the five Berea sandstone cores exhibit substantial overlap during the initial crude oil saturation stage, indicating excellent core homogeneity under identical fluid conditions. This confirms that the intrinsic physical properties of the cores had minimal impact on the experimental outcomes.
As the $\text{CO}_{2}$ flooding process advances, the Berea-1 core—under miscible conditions—consistently shows higher recovery efficiency compared to the immiscible cores. However, with continued $\text{CO}_{2}$ injection at large PV volumes, the recovery performance of the immiscible cores gradually converges and eventually becomes nearly identical, emphasizing the importance of extended $\text{CO}_{2}$ exposure.
Furthermore, the ultimate recovery achieved by miscible flooding exceeds that of the immiscible cases, highlighting both the advantage and necessity of achieving miscible displacement conditions to maximize oil recovery in low-permeability reservoirs.

\subsection{Experimental study on carbon dioxide flooding mechanisms using 2D NMR}\label{sec:II-2}

As described in Sec. \ref{sec:II-0}, to more clearly illustrate the experimental findings, two-dimensional nuclear magnetic resonance (2D NMR) measurements were performed.

\begin{figure}[!htbp]
\centering
\includegraphics[scale=0.08]{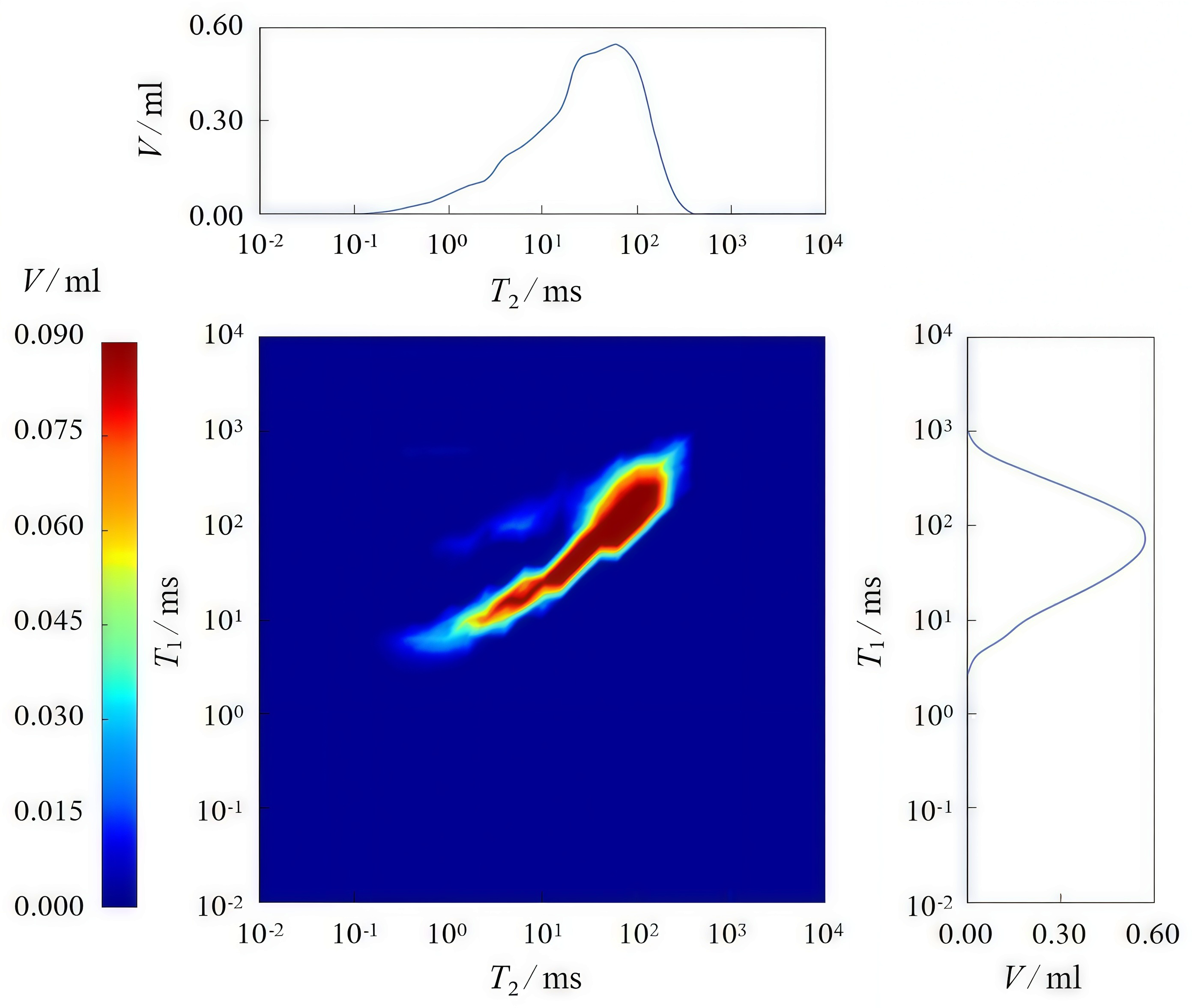}\\
\includegraphics[scale=0.08]{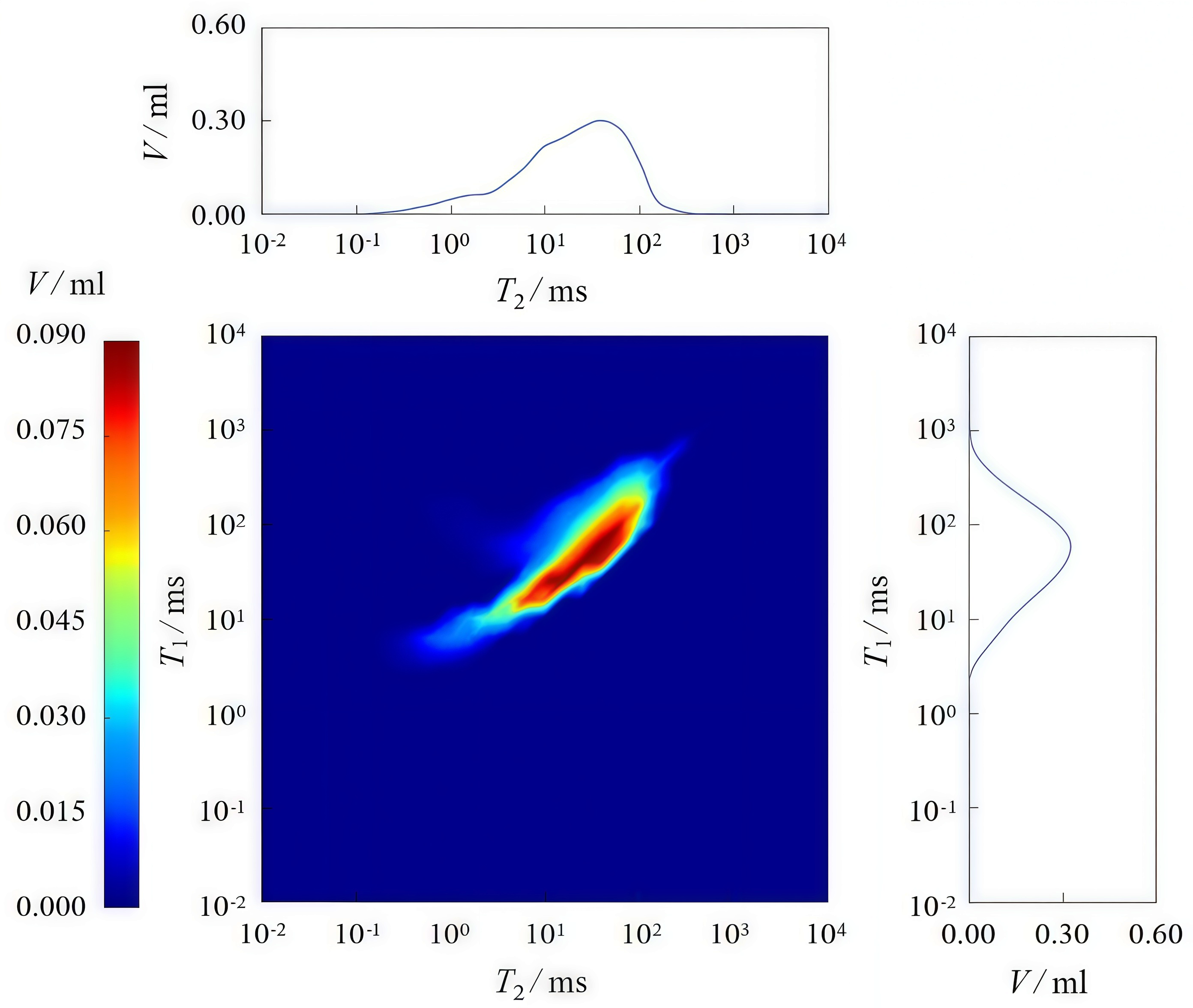}\\
\includegraphics[scale=0.08]{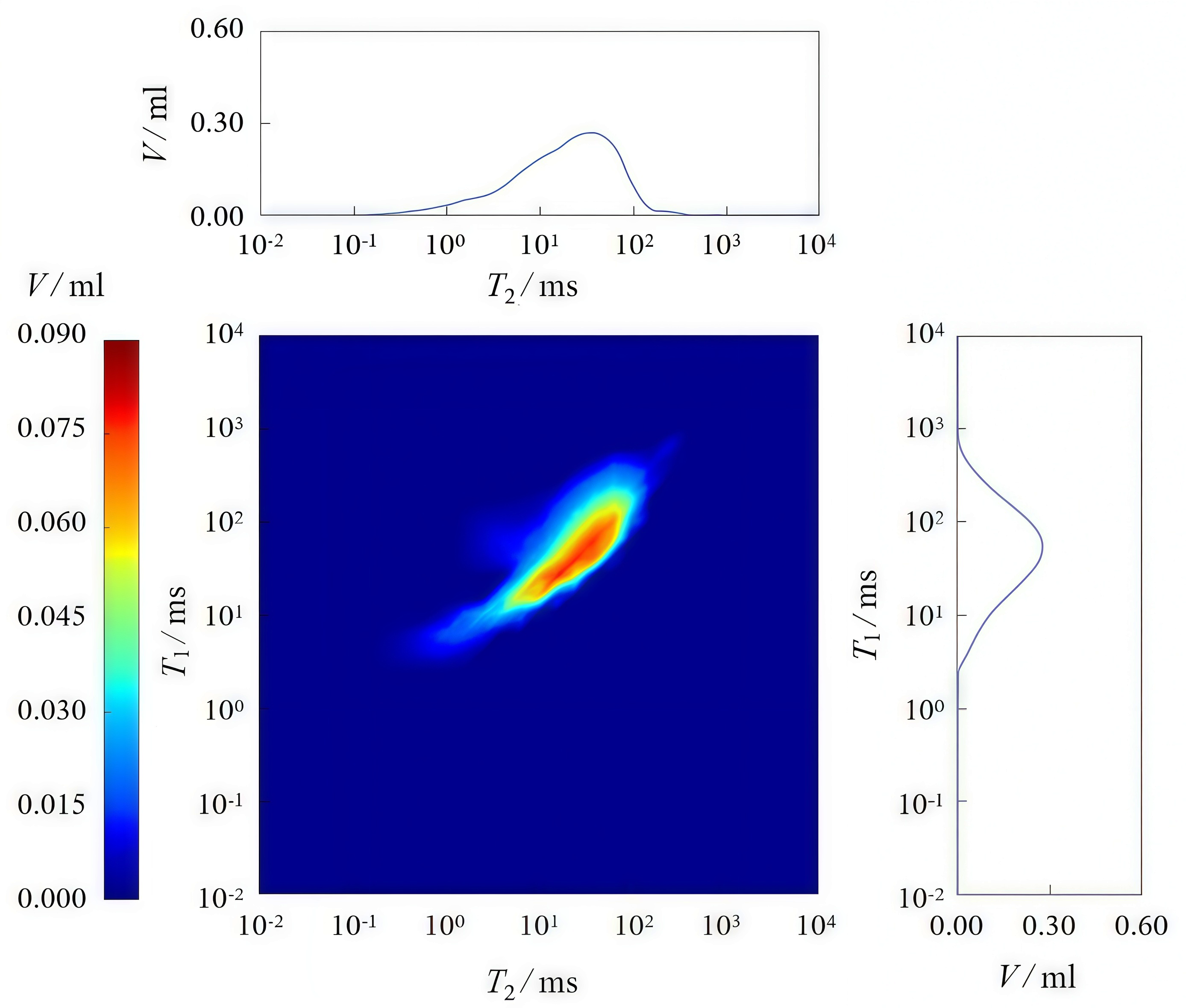}\\
\includegraphics[scale=0.08]{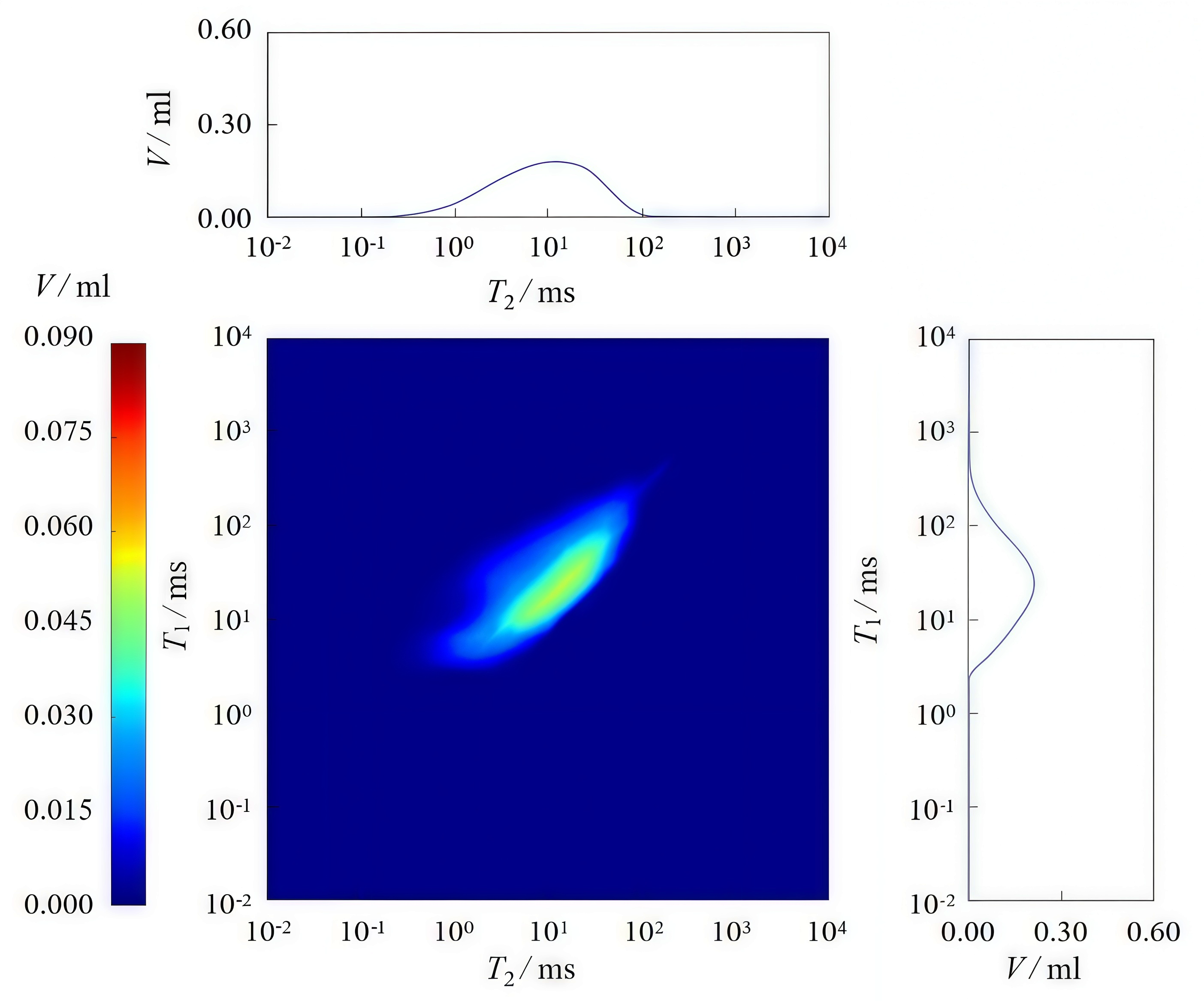}
\caption{Comparison of $T_{1}-T_{2}$ spectra at different stages of $\text{CO}_{2}$ flooding for the Berea-1 core.
 Compared to $T_2$ spectra, $T_{1}-T_{2}$ spectra offer more comprehensive insight into the behavior of different crude oil components during flooding. In this study, the evolution of spectral features highlights the progressive displacement of lighter hydrocarbons. In the $T_{1}-T_{2}$ maps, lighter molecular weight components appear in the upper-right region and are preferentially displaced as CO$_2$ injection proceeds.}
\label{Fig2}
\end{figure}

\begin{figure}[!htbp]
\centering
\includegraphics[scale=0.08]{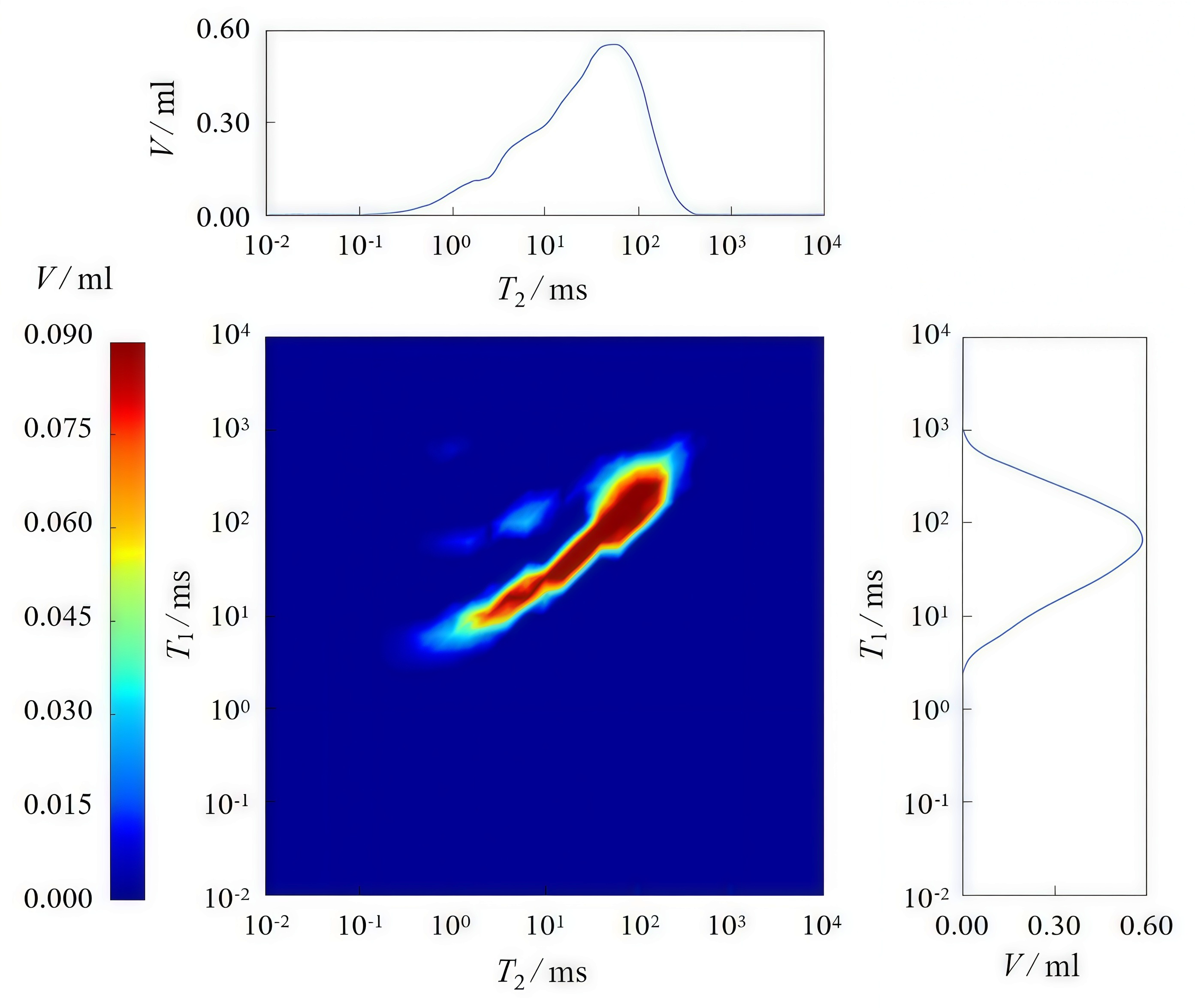}\\
\includegraphics[scale=0.08]{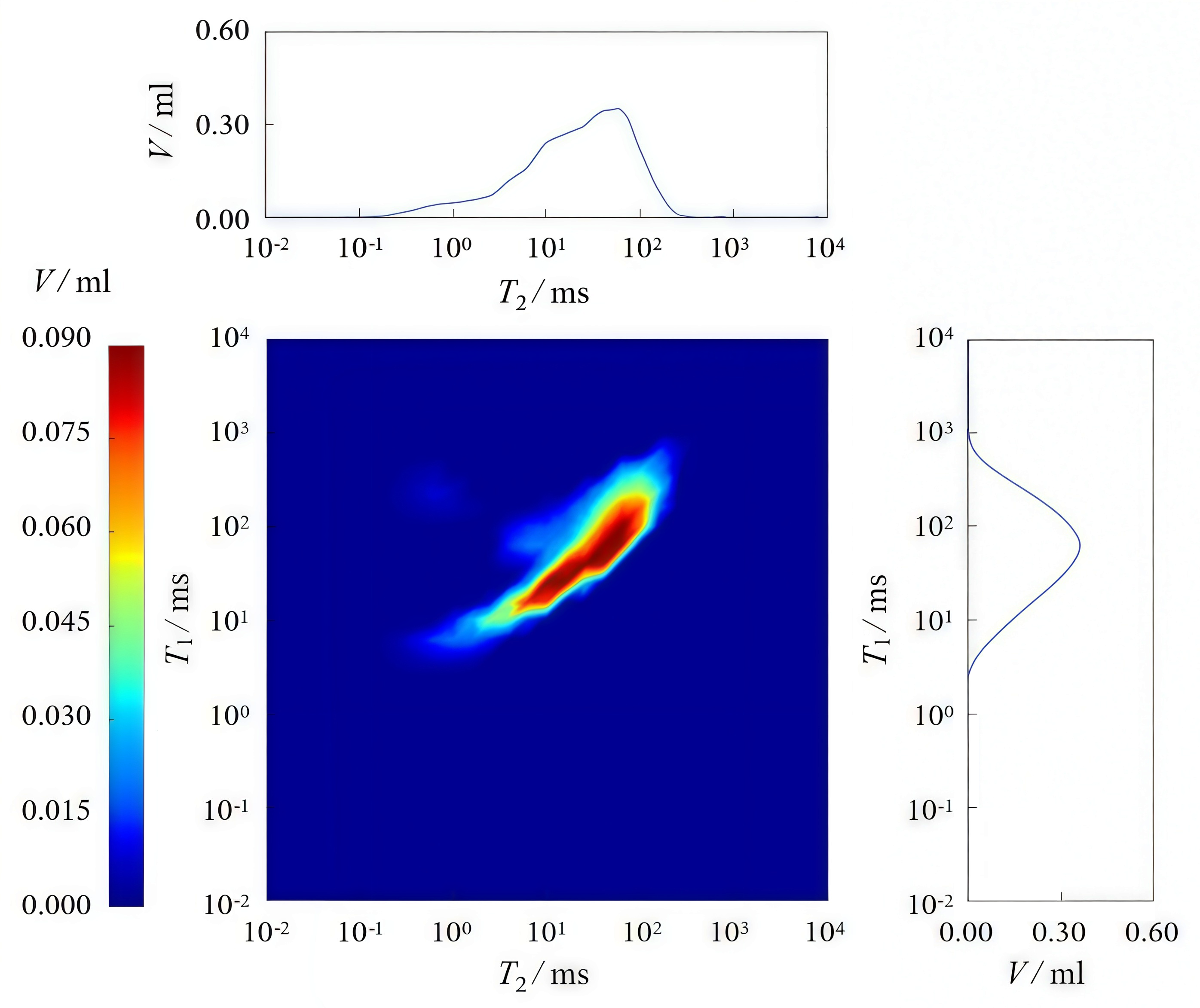}\\
\includegraphics[scale=0.08]{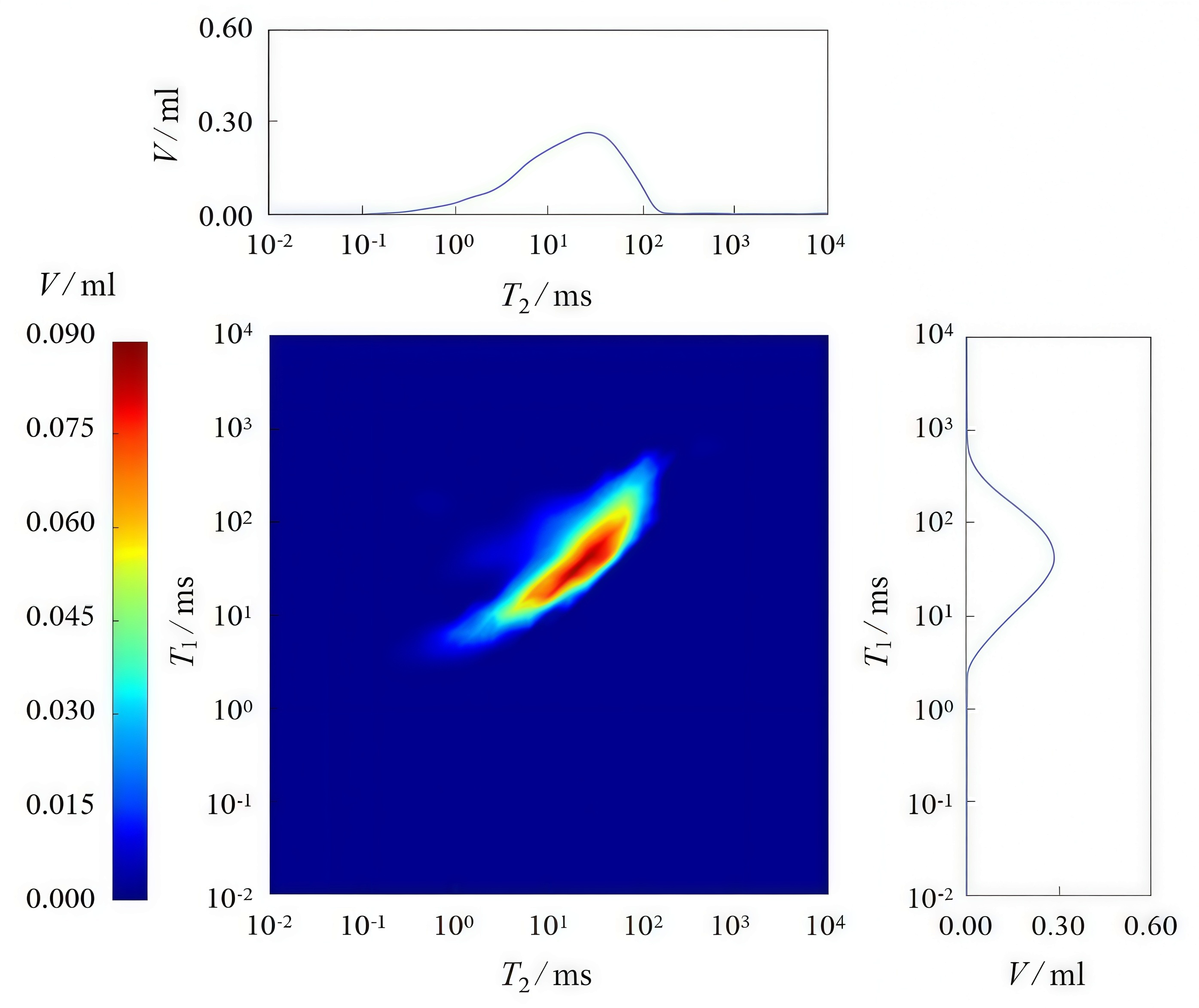}\\
\includegraphics[scale=0.08]{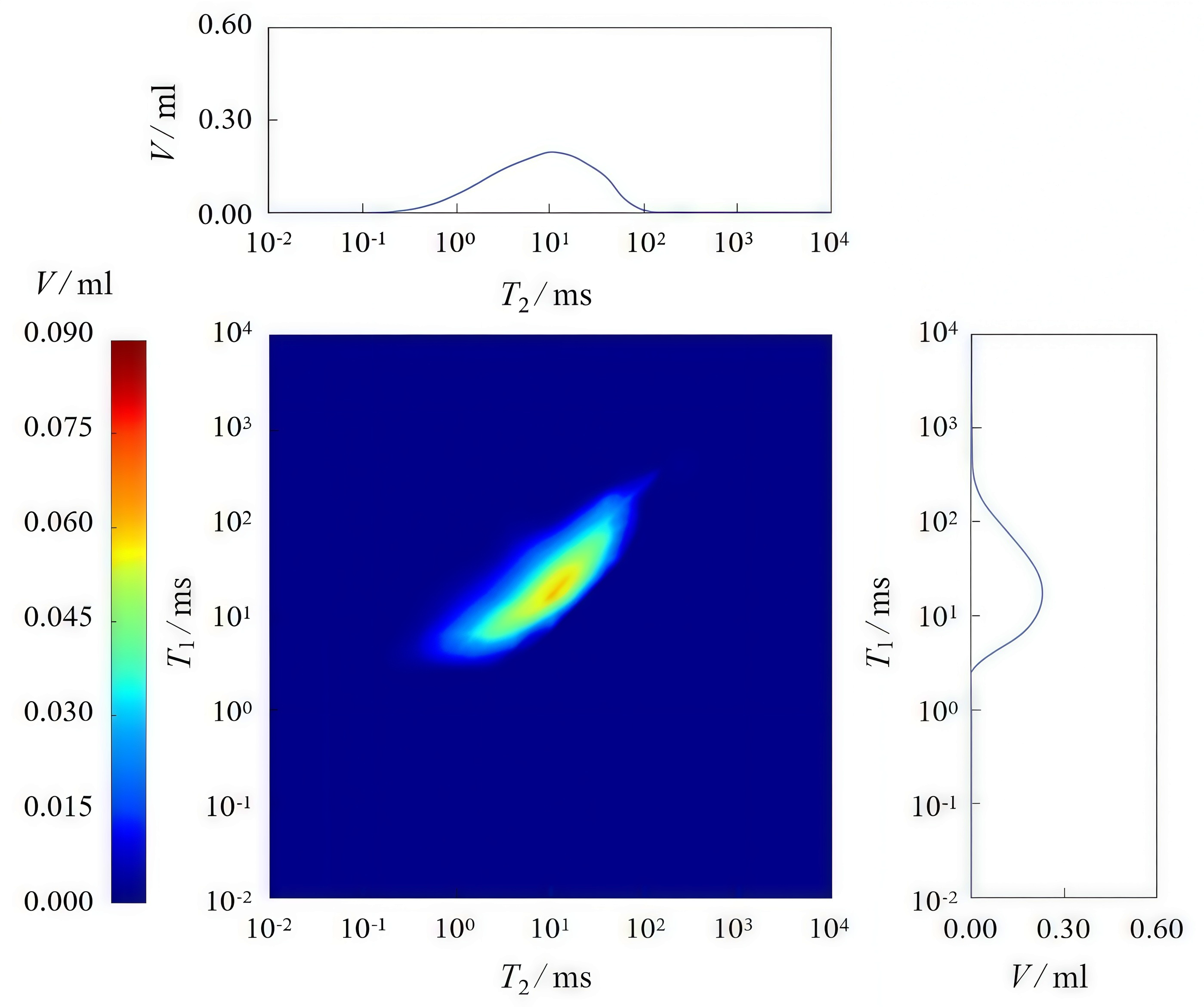}
\caption{Comparison of $T_{1}-T_{2}$ spectra at different stages of $\text{CO}_{2}$ flooding for the Berea-3 core. Berea-3 exhibits a similar spectral evolution trend to Berea-1, with enhanced displacement of lighter components, but retains substantially more total oil.}
\label{Fig3}
\end{figure}

Fig.~\ref{Fig2} and~\ref{Fig3} display representative $T_{1}-T_{2}$ relaxation maps of two Berea sandstone cores, Berea-1 and Berea-3, at different stages of the flooding process.
Berea-1 was subjected to miscible $\text{CO}_{2}$ flooding, whereas Berea-3 underwent immiscible $\text{CO}_{2}$ displacement. These spectra provide insight into fluid distribution evolution under different flooding mechanisms.

At the initial displacement stage (top-left spectrum in each group), free oil accounted for the largest proportion. As $\text{CO}_{2}$ flooding progressed, the 2D spectra evolved through the top-right (after $0.2$ PV), bottom-left (after $0.4$ PV), and bottom-right (after $2.0$ PV) spectra, showing a clear left shift in the transverse relaxation time ($T_{2}$) spectrum and a downward shift in the longitudinal relaxation time ($T_{1}$).
The $T_{1}-T_{2}$ 2D NMR spectra reveal that as the displacement progressed, the fluid content in the Berea cores gradually decreased. The proportion of free oil reduced, while the relative proportion of adsorbed oil increased, with heavy components accumulating. This trend was consistent with the changes observed in the 1D NMR $T_{2}$ spectra.
For the immiscible $\text{CO}_{2}$-flooded Berea-3 core, under insufficient $\text{CO}_{2}$ injection, cores subjected to higher displacement pressures exhibited higher recovery efficiency. This indicates that in the early stages of immiscible $\text{CO}_{2}$ flooding, higher displacement pressure results in higher recovery under the same PV injection. In contrast, the miscible $\text{CO}_{2}$-flooded Berea-1 core yielded higher recovery than the immiscible $\text{CO}_{2}$-flooded cores.

For $\text{CO}_{2}$ flooding, increasing immiscible $\text{CO}_{2}$ displacement pressure can improve recovery efficiency, while miscible $\text{CO}_{2}$ flooding yields the best performance. At higher PV injection volumes, immiscible $\text{CO}_{2}$ flooding exhibits a clear contrast in efficiency compared to miscible $\text{CO}_{2}$ flooding. Although increasing immiscible $\text{CO}_{2}$ displacement pressure improves recovery efficiency, it does not significantly enhance the flowability of heavy crude oil components in the reservoir. In contrast, miscible $\text{CO}_{2}$ flooding can markedly improve the flowability of heavy oil components, leading to superior oil recovery performance.

\section{Theoretical Derivation of Seepage Processes Incorporating Diffusion Effects}\label{sec:III}

\begin{figure}[!htbp]
\centering
\includegraphics[scale=0.38]{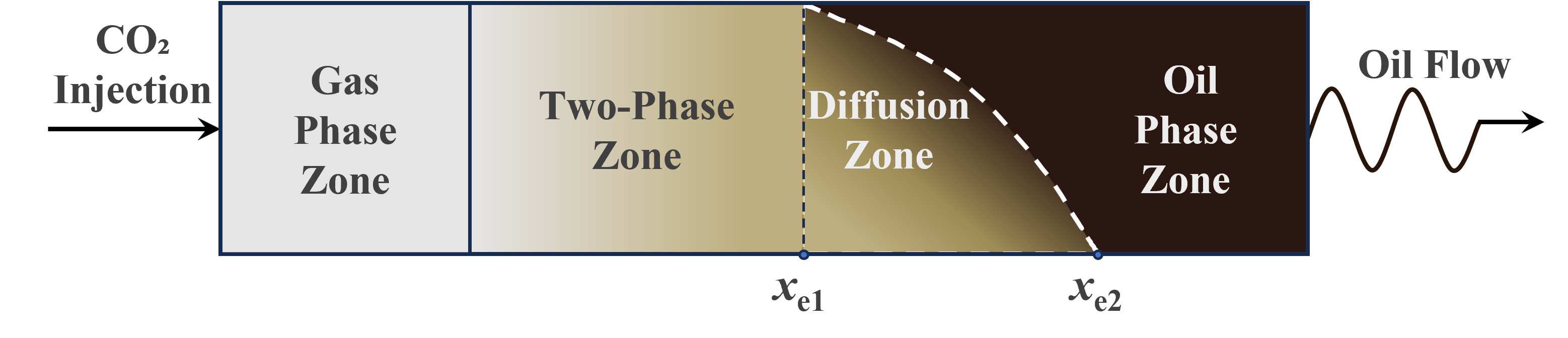}
\caption{Schematic diagram illustrating the multi-phase seepage process during $\text{CO}_{2}$ flooding, showing the interaction between the $\text{CO}_{2}$ gas phase, oil phase, and water phase, as well as the displacement front dynamics.}
\label{Fig4}
\end{figure}

To further investigate the advantages of miscible $\text{CO}_{2}$ flooding, a simplified flow model incorporating diffusion effects is proposed.
Currently, most  studies on $\text{CO}_{2}$ flooding focus primarily on displacement mechanisms and oil recovery efficiency. However, the theoretical modeling of $\text{CO}_{2}$ development often continues to reference analytical frameworks developed for water flooding, which may not accurately represent gas–oil interactions.
Unlike water flooding, gaseous $\text{CO}_{2}$ interacts with reservoir fluids through three key mechanisms: dissolution, diffusion, and convection. These interactions, particularly at the displacement front, lead to fundamental differences in flow behavior.
To account for these effects, we establish a one-dimensional homogeneous porous media model for two-phase $\text{CO}_{2}$–oil flow, as illustrated in Fig.~\ref{Fig4}. The model operates under the following assumptions:
(1) Thermodynamic equilibrium is maintained throughout the process;
(2) Porosity and permeability remain constant;
(3) Viscosity reduction and compressibility effects of $\text{CO}_{2}$ are neglected.
Under these conditions, the reservoir domain is conceptually divided into four distinct zones during the $\text{CO}_{2}$ flooding process:
(a) A pure $\text{CO}_{2}$ gas zone;
(b) A two-phase $\text{CO}_{2}$–oil coexistence zone;
(c) A $\text{CO}_{2}$ diffusion-dominated transition zone;
(d) A pure crude oil zone.

To determine the position of the gas front \( x_{\mathrm{e1}} \), the Buckley–Leverett (B–L) equation,\cite{30} based on the iso-saturation front assumption, is employed.
The B–L equation is given by:
\begin{equation}
x_{e1} - x_0 = \frac{f'_g(S_g)}{\phi A} \int_{0}^{t} q(t) \, dt, \label{eq:BL-1}
\end{equation}
where \( f'_g(S_g) \) is the derivative of the gas fractional flow function, \( \phi \) is porosity, \( A \) is the cross-sectional area, and \( q(t) \) is the volumetric injection rate.
Assuming \( Q = \int_0^t q(t) \, dt \) and \( x_0 = 0 \), Eq. (\ref{eq:BL-1}) becomes:
\begin{equation}
x_{e1} = \frac{f'_g(S_g)}{\phi A} Q. \label{eq:BL-2}
\end{equation}
The gas fractional flow function \( f_g(S_g) \), defined through the relative permeability–saturation relationship, establishes a known correlation between gas saturation and fractional flow. Thus, the gas front location \( x_{\mathrm{e1}} \) can be considered a determinable quantity under given flow and saturation conditions.

The transport of $\text{CO}_{2}$ components during diffusion can be described by the mass conservation equation:
\begin{equation}
\frac{\partial c}{\partial t} + \bm{\nabla} \cdot (\mathbf{v} c) = D \nabla^2 c, \label{eq:CM}
\end{equation}
where \( c \) is the component concentration, $\mathbf{v}$ is the Darcy velocity, and \( D \) is the diffusion coefficient. This equation assumes no source or sink terms, constant \( D \), and incompressible flow. The second term on the left-hand side represents convection, while the right-hand side represents diffusion, based on Fick’s second law. Equation (\ref{eq:CM}) is thus referred to as the convection–diffusion equation.

For one-dimensional flow, the equation simplifies to:
\begin{equation}
\phi \frac{\partial c}{\partial t} + v \frac{\partial c}{\partial x} = D \phi \frac{\partial^2 c}{\partial x^2}, \label{eq:1D}
\end{equation}
with the initial and boundary conditions:
\[
\begin{cases}
c(x_{e1}, t) = c_0, & t \geq 0, \\
c(x, t=0) = 0, & x > x_{e1}, \\
c(x \rightarrow x_{e2}, t) \rightarrow 0, & t \geq 0.
\end{cases}
\]
To generalize, the following dimensionless variables are introduced:
\[
t_Q = \int_0^t \frac{v}{\phi L} \, dt, \quad x_Q = \frac{x}{L}, \quad c_Q = \frac{c}{c_0}, \quad K = \frac{v L}{\phi D},
\]
where $L$ denotes the total investigated depth of the zone, and $K$ is a normalization parameter.
Substituting into Eq. (~\ref{eq:1D}) yields the dimensionless form:
\begin{equation}
\frac{1}{K} \frac{\partial^2 c_Q}{\partial x_Q^2} - \frac{\partial c_Q}{\partial x_Q} = \frac{\partial c_Q}{\partial t_Q}, \label{eq:dimless}
\end{equation}
with boundary conditions:
\[
\begin{cases}
c_Q(x_Q = x_{e1Q}, t_Q) = 1, & t_Q \geq 0, \\
c_Q(x_Q > x_{e1Q}, t_Q = 0) = 0, \\
c_Q(x_Q \rightarrow x_{e2Q}, t_Q) \rightarrow 0, & t_Q \geq 0.
\end{cases}
\]

Applying the Laplace transform to Eq. ~\ref{eq:dimless} and solving with characteristic roots yields:
\begin{equation}
\bar{c}_Q(x_Q, s) = \frac{1}{s} \exp\left[ \frac{(x_Q - x_{e1Q})(1 - \sqrt{1 + 4s/K})}{2/K} \right] \label{eq:laplace}
\end{equation}

Taking the inverse Laplace transform gives the time-domain solution:
\begin{align}
c_{Q}\left(x_{Q},t_{Q}\right) &= \exp\left(K\frac{x_{Q}-x_{e1Q}}{2}\right)\left\{\mathrm{erfc}\left(\frac{x_{Q}-x_{e1Q}+Kt_{Q}}{2\sqrt{Kt_{Q}}}\right)\right.\nonumber\\
&\hspace{-1.2cm}+\exp\left[K\left(x_{Q}-x_{e1Q}\right)\right]\mathrm{erfc}\left(\frac{x_{Q}-x_{e1Q}-Kt_{Q}}{2\sqrt{Kt_{Q}}}\right)\Bigg\}. \label{eq:time solution}
\end{align}
This solution is applicable for arbitrary values of \( K \). Here, \( \mathrm{erfc}(x) \) denotes the complementary error function.

Equation (\ref{eq:time solution}) enables calculation of the compositional front during $\text{CO}_{2}$ flooding with diffusion. Given a known diffusion coefficient and an initial concentration \( c_0 \) at the saturation front \( x_{e1} \), the extent of $\text{CO}_{2}$ diffusion can be determined. The location where concentration approaches zero is defined as the diffusion front \( x_{e2} \).

\begin{figure}[!htbp]
\centering
\includegraphics[scale=0.10]{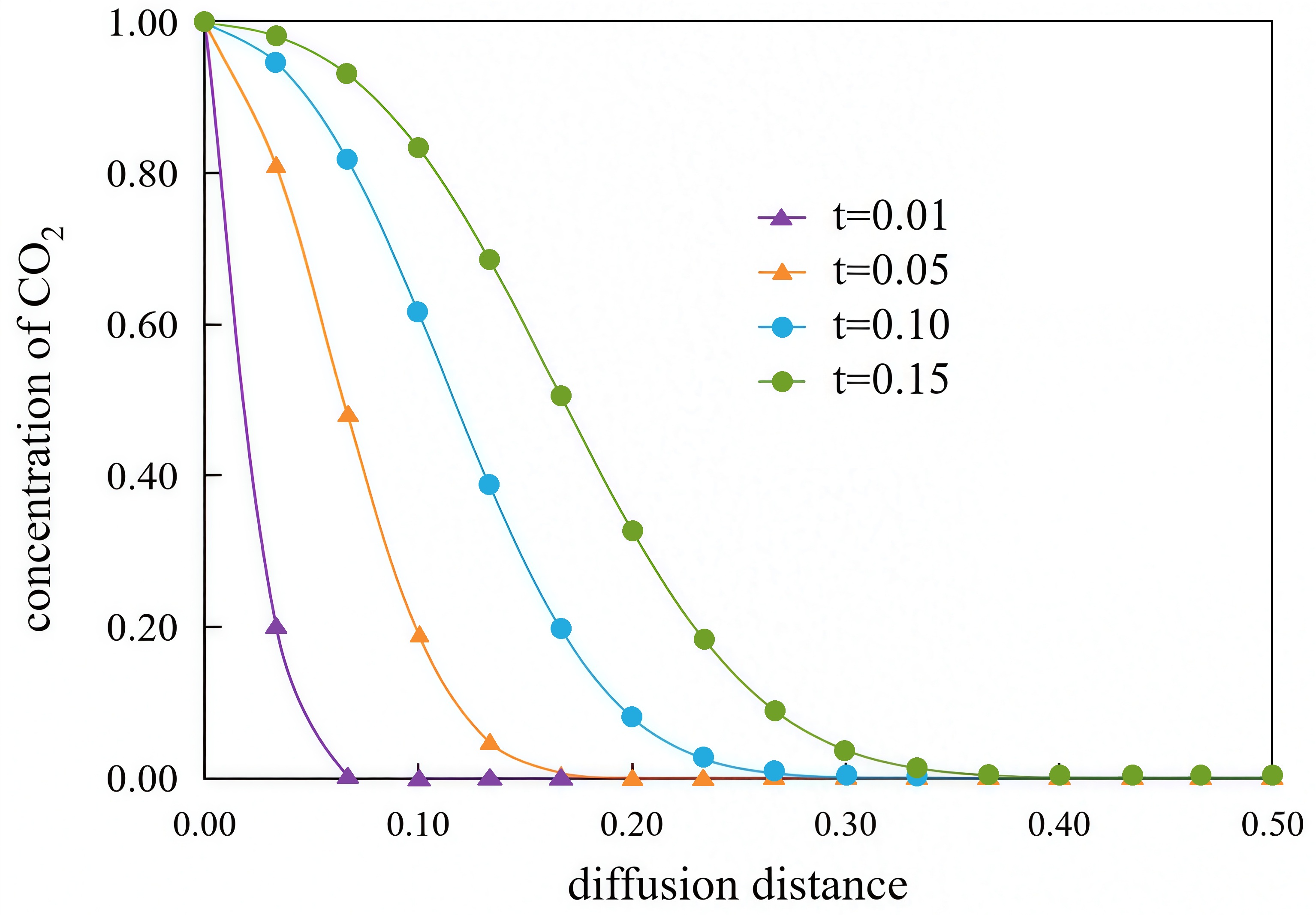}
\caption{$\text{CO}_{2}$ concentration profile over time.}
\label{Fig5}
\end{figure}

Figure~\ref{Fig5} shows the evolution of the $\text{CO}_{2}$ concentration profile over time, calculated using Eq. (\ref{eq:time solution}) with a diffusion coefficient of \( D = 10^{-4} \).
As shown in Fig.~\ref{Fig2}, with the increase in diffusion time, the compositional front due to diffusion gradually shifts forward, expanding the $\text{CO}_{2}$ displacement front. Meanwhile, the concentration of $\text{CO}_{2}$ within the already diffused region also increases. The intersection of the curve with the horizontal axis can effectively predict the time and distance at which the $\text{CO}_{2}$ breakthrough occurs during the displacement process.
Furthermore, this illustrates that under miscible conditions, $\text{CO}_{2}$ rapidly increases the contact area with crude oil, enhancing mass transfer efficiency. This enables more effective extraction of light components, thereby accelerating the overall displacement process. These results are consistent with the experimental findings for Berea-1 core, which, upon entering the miscible flooding stage, exhibited superior recovery compared to other core samples.

\section{Material Balance Equation for Carbon Dioxide Flooding}\label{sec:IV}

Experimental studies often face inherent limitations, particularly under extreme conditions such as ultra-low permeability reservoirs, high-pressure environments, and geologically complex systems (e.g., heterogeneous formations). Additionally, experimental results are frequently constrained by equipment resolution, core heterogeneity, and limited control over boundary conditions, which hinders the ability to fully elucidate the individual contributions of mechanisms involved in $\text{CO}_{2}$ flooding.

To address these limitations, this study proposes a generalized and scalable mathematical model for $\text{CO}_{2}$ displacement that can quantitatively describe the coupled effects of key physical mechanisms—including diffusion, extraction, and dissolution—under a range of reservoir conditions. The model is built upon the reservoir material balance equation, which serves as the theoretical foundation. By calibrating the model against existing experimental data, a complementary relationship is established, enabling mutual verification between simulation and laboratory observations. This approach compensates for the limited experimental capacity under extreme conditions and improves overall understanding of the displacement process.
Furthermore, the proposed model offers predictive capabilities for identifying potential operational risks, such as exceeding formation fracture pressure or early $\text{CO}_{2}$ breakthrough. It provides theoretical guidance for experimental parameter selection and field-scale injection design, helping to avoid high-risk and low-efficiency operating conditions while optimizing oil recovery performance.

In previous studies, material balance equations for reservoirs using traditional water flooding or $N_{2}$ displacement have been well-established. However, there are significant differences between the displacement models for $\text{CO}_{2}$ flooding and those for water flooding or $N_{2}$ displacement. In gas displacement processes, the gas flow rate is typically approximated using an average pressure, which causes substantial differences between the gas and water flooding models. By applying material conservation principles to establish reservoir material balance equations, we can avoid the complexities of analyzing phase-specific displacement mechanisms. This approach allows for a more comprehensive evaluation of overall reservoir development, focusing on macro-scale strategies for efficient oil recovery.

Typically, it is assumed that the displacing fluid does not interact with any reservoir components, including formation crude oil, irreducible water, and the rock matrix. This assumption is reasonable for water flooding or $N_{2}$ flooding processes. However, such a completely non-reactive two-phase flow system does not apply to $\text{CO}_{2}$ flooding. $\text{CO}_{2}$ flooding is inherently a reactive process, where $\text{CO}_{2}$ interacts with formation crude oil through mutual mass transfer and diffusive exchange. Additionally, $\text{CO}_{2}$ reacts with the rock matrix, leading to natural formation consumption as $\text{CO}_{2}$ is absorbed by the reservoir. While this process represents an environmentally favorable $\text{CO}_{2}$ sequestration mechanism, it impacts material balance calculations for $\text{CO}_{2}$ flooding.
From a production standpoint, the material balance equation for $\text{CO}_{2}$ flooding must account for both produced gas and oil. To simplify the reservoir's physical model, irreducible water effects are neglected, and water-free production is assumed. Consequently, the equation considers: (a) the $\text{CO}_{2}$ depletion in the produced gas, and (b) the dissolved $\text{CO}_{2}$ remaining in the crude oil. These characteristics are distinctive in the material balance formulation for $\text{CO}_{2}$ flooding reservoirs.

The reservoir material balance equation is essentially a mass conservation equation. The following assumptions are made for the reservoir model:

(1) The reservoir exhibits excellent homogeneity, with essentially uniform physical properties of both the rock matrix and contained fluids, satisfying isotropic conditions.

(2) Stress transmission within the reservoir is instantaneous and sensitive. At any given time $t$, formation pressure remains consistent throughout the reservoir, with the model maintaining stable mechanical equilibrium.

(3) The entire reservoir development system is treated as an isolated mass system with no mass exchange with external environments. However, the system possesses ideal heat transfer characteristics, maintaining a constant background temperature $T_{0}$ throughout development.

(4) The following effects are neglected: influence of irreducible water and water influx; gravitational effects during development; capillary pressure impacts on reservoir production; gas slippage effects; initial solution gas presence; and potential reservoir compaction during production.

(5) Oil production rates remain balanced across all reservoir regions during development.

Produced fluids are replaced by internal system expansion and injected fluids; otherwise, a vacuum would form in the reservoir, which contradicts physical principles. In this study, total oil and gas production equals the expansion volume of crude oil and rock, plus the volume of injected gas, yielding the $\text{CO}_{2}$ flooding material balance equation
\begin{align}
\frac{N_{p}B_{o}}{\gamma}&+N_{p}\left(R_{p}+\beta\right)B_{g} = N\left(B_{o}-B_{oi}\right)+NB_{oi}C_{f}\Delta p+G_{i}B_{g}\nonumber\\
&+ \left[\left(1-\alpha\right)G_{i}-N_{p}\left(R_{p}+\beta\right)B_{g}\right]\left(B_{g}-B_{gi}\right).\label{eq:MB equation}
\end{align}

All the parameters in Eq. (\ref{eq:MB equation}) will be illustrated in the following text. Eq. (\ref{eq:MB equation}) shows that the left-hand side is composed of two terms, which represent oil production and gas production, respectively. Over a specified development period \( t \), the production volume consists of the oil phase as \( N_p \) and the gas phase as \( G_p \). The effect of irreducible water is neglected, assuming no water production. The dissolution of  $\text{CO}_{2}$ in the \( N_p \) oil phase causes oil phase expansion, with the expansion coefficient \( \gamma \). At the current formation pressure \( p \), the measurable volume factor \( B_o \) accounts for the volume differences of crude oil under varying pressures. Thus, the cumulative volume of \( N_p \) oil production at pressure \( p \) is \( N_p B_o / \gamma \), corresponding to the first term on the left-hand side of Eq. (\ref{eq:MB equation}).
The overall gas-oil ratio, defined as \( R_p = G_p / N_p \), is measurable. Under the assumption of no initial dissolved gas, \( G_p \) is considered to consist entirely of the displacing gas  $\text{CO}_{2}$, which replaces part of the dissolved gas. This contribution is represented by the second term on the left-hand side of Eq. (\ref{eq:MB equation}). Both the dissolution ratio \( \beta \) and the gas phase volume factor of $\text{CO}_{2}$, \( B_g \), can be determined from laboratory measurements of the crude oil’s properties.

In Eq. (\ref{eq:MB equation}), the right-hand side consists of four terms: crude oil expansion, rock matrix expansion, injected gas volume, and the expansion effect due to the pseudo-dissolution gas drive formed by $\text{CO}_{2}$. \( N \) represents the total crude oil volume under surface conditions (original oil in place). \( B_{oi} \) denotes the oil formation volume factor at the initial reservoir pressure \( p_{i} \), and \( N(B_o - B_{oi}) \) quantifies the oil expansion resulting from fluid production, corresponding to the first term on the right-hand side of Eq. (\ref{eq:MB equation}). As production time \( t \) progresses, reservoir pressure decreases from the initial pressure \( p_{i} \) to the final pressure \( p \), with the pressure drop \( \Delta p = p_{i} - p \) being measurable. The model neglects the effects of initial gas caps and solution gas, so despite the pressure drop \( \Delta p \), there is no expansion of the gas cap or the rock matrix in the gas cap region. The model also ignores irreducible water and natural water influx, assuming a pure $\text{CO}_{2}$ flooding process with no water injection, and thus no aqueous phase volume changes. In the absence of irreducible water, the pore volume is equal to the initial oil volume, \( V_p = N B_{oi} \). By incorporating measurable formation compressibility \( C_f \) , the rock expansion volume is calculated as \( V_p C_f \Delta p = N B_{oi} C_f \Delta p \), corresponding to the second term on the right-hand side of Eq. (\ref{eq:MB equation}). The cumulative injected $\text{CO}_{2}$ volume is \( G_i B_g \), where \( G_i \) is the surface-volume of $\text{CO}_{2}$, representing the third term on the right-hand side of Eq. (\ref{eq:MB equation}). The $\text{CO}_{2}$ flooding process induces a pseudo-dissolution gas drive effect, leading to oil expansion due to the $\text{CO}_{2}$ retained in the reservoir. This expansion combines the oil and $\text{CO}_{2}$ volumetric expansions for the drive mechanism. It is important to note that formation-absorbed $\text{CO}_{2}$ contributes solely to storage. Therefore, the expansion term becomes \( \left[\left(1 - \alpha \right) G_i - N_p \left(R_p + \beta \right) B_g \right]\left(B_g - B_{gi}\right) \), where \( \alpha \) is the $\text{CO}_{2}$ absorption coefficient and \( B_{gi} \) is the $\text{CO}_{2}$ formation volume factor at the initial pressure \( p_i \), representing the supercritical $\text{CO}_{2}$ volume factor under initial conditions. This corresponds to the fourth term on the right-hand side of Eq. (\ref{eq:MB equation}).

From Eq. (\ref{eq:MB equation}), the total crude oil volume $N$ could be derived as:
\begin{align}
N&=\frac{1}{B_{o}-B_{oi}+B_{oi}C_{f}\Delta p} \Bigg\{N_{p}B_{o}/\gamma +N_{p}\left(R_{p}+\beta\right)B_{g}-G_{i}B_{g} \nonumber\\
&\quad - \left[\left(1-\alpha\right)G_{i}-N_{p}\left(R_{p}+\beta\right)B_{g}\right]\left(B_{g}-B_{gi}\right)\Bigg\}.\label{eq:total crude oil volume}
\end{align}

All parameters on the right-hand side of Eq. (\ref{eq:total crude oil volume}) can be determined through laboratory analysis and field measurements. Additionally, Eq. (\ref{eq:MB equation}) allows for cross-verification of the remaining parameters by comparing them with estimated reservoir reserves.

\begin{figure}[!htbp]
\centering
\includegraphics[scale=0.12]{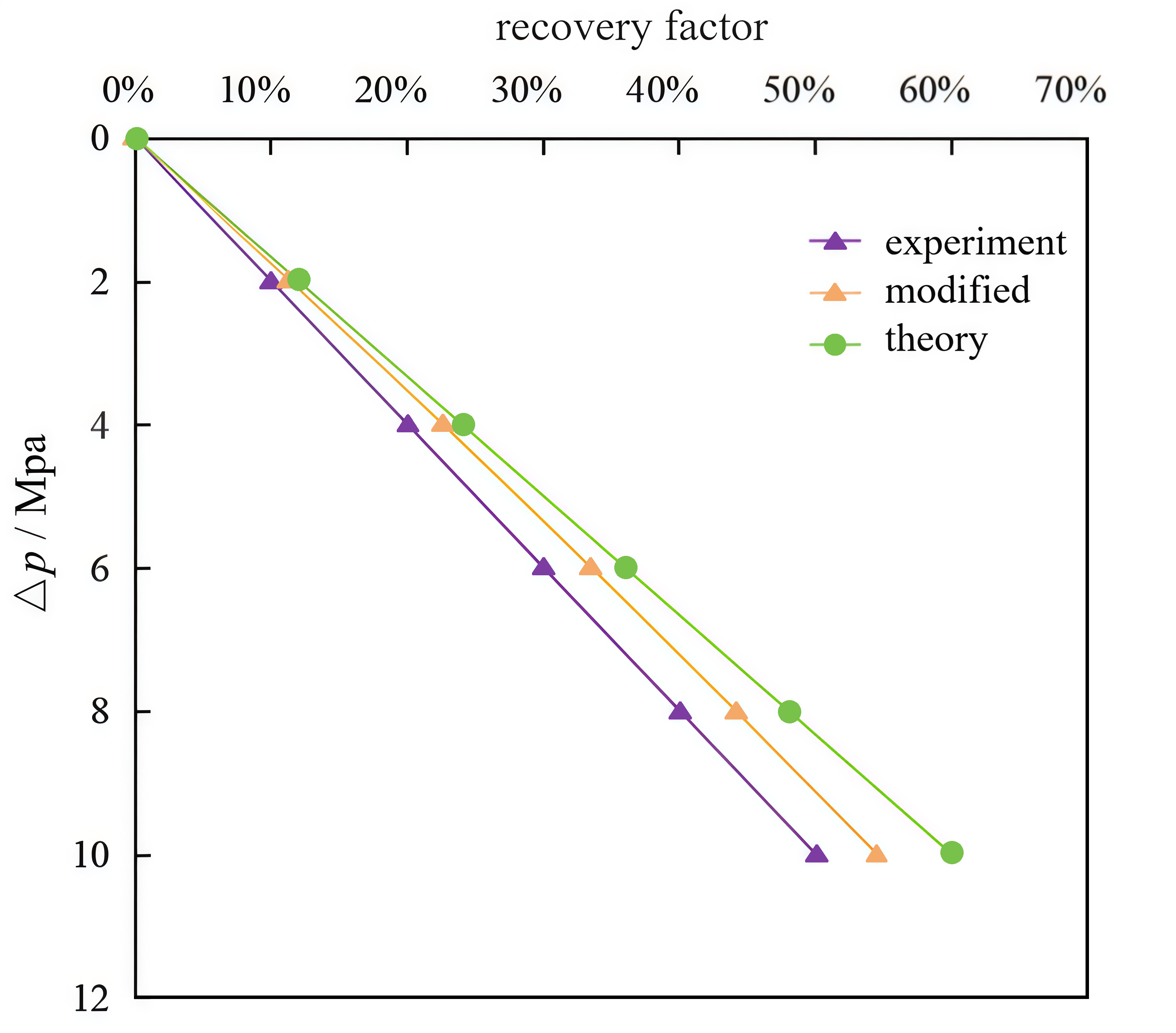}
\caption{Relationship between recovery factor and reservoir pressure drop.}
\label{Fig6}
\end{figure}

Based on the actual production data from Section \ref{sec:II}, the $\text{CO}_{2}$ flooding reservoir material balance equation is used to calculate the modified production data. The traditional reservoir material balance equation is employed to calculate theoretical production data. A comparison is made between the actual production, the modified production, and the theoretical production, and a relationship curve between the recovery factor and the reservoir pressure drop is plotted in Fig. \ref{Fig6}.

From Fig. \ref{Fig6}, it is evident that for the same pressure drop \(\Delta p\), the actual production is lower than the theoretical production of the reservoir. The modified production, however, lies between the actual and theoretical production, indicating that the  $\text{CO}_{2}$ flooding reservoir material balance equation serves to adjust the material balance of the reservoir. The  $\text{CO}_{2}$ flooding reservoir material balance equation is therefore more applicable to reservoirs under  $\text{CO}_{2}$ flooding development.

The modified production matches the theoretical production calculated using the traditional material balance equation in the early stages of the displacement process. As the displacement progresses, and the reservoir pressure decreases, the pressure drop \(\Delta p\) increases, causing the modified production to deviate more significantly from the theoretical production. This suggests that the $\text{CO}_{2}$ flooding reservoir material balance equation is more accurate for reservoirs with higher production in $\text{CO}_{2}$ flooding development, compared to the traditional reservoir material balance equation.

\section{Conclusions and Discussions}\label{sec:V}

This study systematically investigated the mechanisms and efficiency of $\text{CO}_{2}$ flooding in low-permeability reservoirs through comprehensive core experiments and theoretical modeling. The main findings are summarized as follows:

(1) Enhanced recovery performance of $\text{CO}_{2}$ flooding:
The experimental results demonstrate that $\text{CO}_{2}$ flooding is particularly effective for low-permeability, ultra-low permeability, and tight reservoirs. $\text{CO}_{2}$ interacts with crude oil through multiple mechanisms including viscosity reduction, oil swelling, and extraction of light components, modifing the solid-liquid interactions between the formation and crude oil to enhance formation permeability, collectively improving oil mobility and displacement efficiency. This effectively mitigates some challenges in conventional water flooding (e.g., poor mobilization of residual oil) and surfactant-polymer (SP) flooding (e.g., polymer-induced plugging in low-permeability zones).

(2) Superiority of miscible flooding:
The experimental results demonstrate that under miscible conditions (24.00 MPa in this study), the ultimate recovery reached 60.97\%, 3.44\% higher than the maximum recovery of 57.53\% achieved in immiscible flooding, while the differences in the other four core samples did not even exceed 3\%. The comparative analysis between miscible and immiscible flooding reveals that miscible displacement exhibits distinct advantages. The complete dissolution of $\text{CO}_{2}$ into crude oil under miscible conditions creates a pseudo-single phase flow, eliminating interfacial tension and enabling more efficient displacement of both light and heavy oil components. Moreover, $\text{CO}_{2}$ flooding technology is not limited by chemical agent failure and can achieve a higher recovery rate than the theoretical limit during the displacement process. Therefore, $\text{CO}_{2}$ flooding technology is more suitable for light oil reservoirs with low miscible pressure and no reliance on chemical agents to reduce viscosity than medium-heavy oil reservoirs. In practical applications, priority should be given to achieving miscible displacement by increasing the injection pressure to maximize the oil recovery effect.

(3) Optimization of injection parameters:
The PV-dependent recovery behavior provides critical insights for field application strategies. The experimental data show that:
Initial rapid production can be achieved with small PV injections (0.2 PV yielding ~47.40\% recovery in miscible case);
Sustained production requires larger PV injections (2.0 PV increasing recovery to 60.97\%);
The marginal recovery gain decreases with increasing PV, suggesting an economic optimum for field operations.

(4) Diffusion-enhanced displacement mechanism:
The developed convection-diffusion model quantitatively describes the important role of molecular diffusion in $\text{CO}_{2}$ flooding. Numerical solutions of the dimensionless transport equation reveal that:
Diffusion significantly extends the $\text{CO}_{2}$ penetration depth beyond the displacement front;
The diffusion-dominated mass transfer accelerates the establishment of miscible conditions;
The time-dependent concentration profiles enable more accurate prediction of gas breakthrough.

(5) Improved material balance approach:
The modified material balance equation incorporating $\text{CO}_{2}$-crude oil interactions and diffusion effects shows superior predictive capability compared to conventional models. Key improvements include:
Explicit accounting for $\text{CO}_{2}$ dissolution in oil (through parameter $\beta$);
Consideration of $\text{CO}_{2}$ sequestration in rock matrix (parameter $\alpha$);
Better alignment with experimental data, particularly at higher recovery factors.

In summary, $\text{CO}_{2}$ flooding is a highly effective method for developing low-permeability light oil reservoirs, where miscible displacement and large-PV-number injection serve as key strategies for enhancing oil recovery. Future research should further investigate the dynamic behavior of $\text{CO}_{2}$ flooding under extreme conditions, coupled with numerical simulation to optimize field application schemes, thereby enabling more efficient and economical hydrocarbon development.

The primary challenge in $\text{CO}_{2}$ flooding is that reservoir heterogeneity causes $\text{CO}_{2}$ to break through along high-permeability zones, leaving crude oil in low-permeability areas poorly displaced. Research efforts must focus on expanding $\text{CO}_{2}$ sweep efficiency, such as using water-alternating-gas (WAG) injection, viscosity-enhancing agents, or nanoparticles to increase $\text{CO}_{2}$ viscosity, combined with foam flooding and other techniques to block high-permeability channels and improve sweep volume.
Additionally, to address the issue of gas channeling during $\text{CO}_{2}$ flooding, the development of multi-media composite conformance control systems is essential. This includes acid-resistant gel systems, in-situ emulsion-based conformance control systems, and self-adaptive viscosity-enhanced conformance control systems. Further advancements should focus on deep diversion systems with controllable migration and low-cost large-channel blockage systems to enhance $\text{CO}_{2}$ flooding regulation.
Synergistic integration with emerging conformance control technologies should also be prioritized. Research on multi-media composite conformance control systems and novel conformance control agents—such as combining $\text{CO}_{2}$ miscibility with surfactant mobility control—should be evaluated on a case-by-case basis.
From an engineering perspective, optimization of injection and production processes, along with advancements in monitoring technologies, is crucial. A surface-subsurface integrated collaborative flooding system should be established to achieve full-process optimization—from $\text{CO}_{2}$ capture, transportation, and injection to production—maximizing overall efficiency.

\section*{Acknowledgements}
We gratefully acknowledge the time and effort devoted by the Editor and Reviewers in evaluating our manuscript and providing insightful feedback, which has significantly contributed to improving the quality and clarity of this work.
This work was supported by the China National Petroleum Corporation (CNPC) Science and Technology Special Project (Grant No. 2023ZZ0802), the CNPC Forward-Looking Basic Research Project (Grant No. 2025DJ10302), the Hebei Outstanding Youth Science Foundation (Grant No. A2023409003), and the Science Foundation of NCIAE (Grant Nos. ZD-2025-06 and KY-2025-03).

\section*{Data Availability}
The data that support the findings of this study are available from the corresponding author upon reasonable request.

\section*{Nomenclature}

$x_{e1}$\quad position of the gas front, m

$S_{g}$\quad gas saturation, dimensionless

$\phi$\quad porosity, dimensionless

$A$\quad cross-sectional area, m$^{2}$

$q$\quad volumetric injection rate,  m$^{3}$/s

$c$\quad component concentration, \%

$v$\quad Darcy velocity, m/s

$D$\quad diffusion coefficient,  m$^{2}$/s

$L$\quad total investigated depth of the zone, m

$K$\quad normalization parameter, dimensionless

$N_{p}$\quad production volume of the oil phase, m$^{3}$

$G_{p}$\quad production volume of the gas phase, m$^{3}$

$\gamma$\quad expansion coefficient, dimensionless

$B_{o}$\quad factor accounts for the volume differences of oil under varying pressures, dimensionless

$R_{p}$\quad overall gas-oil ratio, dimensionless

$\beta$\quad dissolution ratio, dimensionless

$B_{g}$\quad factor accounts for the volume differences of $\text{CO}_{2}$ under varying pressures, dimensionless

$N$\quad original oil in place, m$^{3}$

$p_{i}$\quad initial reservoir pressure, MPa

$B_{oi}$\quad factor accounts for the volume differences of oil under initial reservoir pressure, dimensionless

$\Delta p$\quad pressure drop, N/m$^{2}$

$C_{f}$\quad formation compressibility, MPa$^{-1}$

$G_{i}$\quad initial surface-volume of $\text{CO}_{2}$, m$^{3}$

$\alpha$\quad $\text{CO}_{2}$ absorption coefficient, dimensionless

$B_{gi}$\quad factor accounts for the volume differences of $\text{CO}_{2}$ under initial reservoir pressure, dimensionless

\section*{References}
\bibliography{final}

\end{document}